\documentclass[aps,twocolumn,superscriptaddress,nofootinbib,a4paper,longbibliography]{revtex4-2}

\usepackage{times}
\usepackage{epsfig}
\usepackage{amsfonts}
\usepackage{amsthm}
\usepackage{amssymb}					
\usepackage{amsthm}
\usepackage{dsfont}
\usepackage{bm}
\usepackage{color}
\usepackage{multirow}
\usepackage[normalem]{ulem}
\newcommand{\stkout}[1]{\ifmmode\text{\sout{\ensuremath{#1}}}\else\sout{#1}\fi}
\usepackage{latexsym}
\usepackage{natbib}
\usepackage{verbatim}
\usepackage[T1]{fontenc}
\usepackage{float}
\usepackage{graphicx}
\usepackage{subcaption}
\usepackage{physics}
\usepackage{soul}
\usepackage{caption}
\usepackage{subcaption}
\usepackage{diagbox}
\usepackage{bbm}
\usepackage{tablefootnote}
\usepackage{adjustbox}
\usepackage[dvipsnames]{xcolor}
\usepackage[font=small,labelfont=bf]{caption}
\usepackage{url}

\usepackage{xspace}

\theoremstyle{definition}

\newcommand{\bracket}[3]{\langle#1|#2|#3\rangle}
\newcommand{\expect}[1]{\langle#1\rangle}

\newcommand{\figref}[1]{Fig.~\ref{#1}}

\newcommand{\tabref}[1]{Tab.~\ref{#1}}

\usepackage{amssymb}
\usepackage[customcolors]{hf-tikz} 
\usetikzlibrary{patterns}
\usetikzlibrary{matrix,decorations.pathreplacing}

\pgfkeys{tikz/mymatrixenv/.style={decoration={brace},every left delimiter/.style={xshift=8pt},every right delimiter/.style={xshift=-8pt}}}
\pgfkeys{tikz/mymatrix/.style={matrix of math nodes,nodes in empty cells,left delimiter={[},right delimiter={]},inner sep=1pt,outer sep=1.5pt,column sep=2pt,row sep=2pt,nodes={minimum width=20pt,minimum height=10pt,anchor=center,inner sep=0pt,outer sep=0pt}}}
\pgfkeys{tikz/mymatrixbrace/.style={decorate,thick}}

\usepackage[colorlinks=true,linkcolor=blue,citecolor=magenta,urlcolor=blue]{hyperref}

\begin{document}
	

\title{Semidefinite block-matrix relaxations for computing quantum correlations}

\author{Nicola D'Alessandro}
\affiliation{Physics Department and NanoLund, Lund University, Box 118, 22100 Lund, Sweden.}
		
\author{Carles Roch i Carceller}	
\affiliation{Physics Department and NanoLund, Lund University, Box 118, 22100 Lund, Sweden.}
	
\author{Armin Tavakoli}
\affiliation{Physics Department and NanoLund, Lund University, Box 118, 22100 Lund, Sweden.}
	
\begin{abstract}
Bounding the correlations predicted by quantum theory is an important challenge in quantum information science. Today's leading approach is semidefinite programming relaxations, but existing methods still cannot account for many relevant types of constraints.  Here, we propose a general semidefinite relaxation methodology that can incorporate a breadth of constraints needed in various quantum correlation problems, thereby generalising the seminal Navascués-Pironio-Ac\'in hierarchy. It yields useful results at reasonable computational cost. We showcase the methodology and its features by using it to address five different quantum information problems. These are (i) entanglement witnessing from imperfect measurement devices, (ii) certifying measurements from fidelity-constrained sources, (iii) computing dimensionality in genuine multi-particle entangled states, (iv) benchmarking dimensionality for state preparation devices, and (v) finding uncertainty relations for nearly anti-commuting observables. These applications reflect both the usefulness and versatility of the methodology, as well as its potential for broader relevance in the field.   
\end{abstract}
	
	\date{\today}

	\maketitle

\section{Introduction}
Quantum  science and technology regularly needs protocols and tests to be optimised over the available quantum resources. Characterising the associated statistical predictions of quantum theory, usually referred to as quantum correlations, is therefore an important  endeavour. Problems of this type are broadly relevant: they encompass for example detecting entanglement properties, benchmarking performance in device-independent quantum information and bounding quantum communication advantages.  Most frontier problems in quantum correlations involve simultaneously determining the optimal use of several different quantum resources, such as states and measurements. This usually makes the problems challenging and analytical solutions rare. Semidefinite programs (SDPs) provide an avenue to meet this challenge, by casting them as linear optimisations over positive semidefinite matrices subject to linear constraints. However, since most state-of-the-art correlations problems are not directly solvable by SDP, the main idea has been to replace the original problem with a sequence of appropriate relaxations, each solvable by SDP, to thereby obtain increasingly accurate approximations to the original problem.

Working most often on a case-to-case basis, SDP relaxations have been developed for a breadth of different quantum correlation problems \cite{SDP_rev_tavakoli}. Some prominent examples include the separability problem \cite{Doherty2004}, contextuality \cite{Acin2015}, steering \cite{Pusey2013}, dimension witnessing  \cite{NavascuesVertesi2015}, self-testing \cite{Yang2014}, network nonlocality \cite{Pozas2019, Wolfe2021}, entanglement-assisted communication \cite{Pauwels2021} and device-independent quantum cryptography \cite{Brown2024}. A large share of modern SDP relaxation techniques for quantum correlations build on the core idea underpinning the trail-blazing method of Navascu\'es-Pironio-Ac\'in (NPA) \cite{Pironio2010} which, in a nutshell, characterises correlations via SDP relaxations formed by inner-products of the  operators appearing in the problem.

NPA-type approaches have been most successful for quantum correlation problems with  simple algebraic structure. The hallmark example is quantum nonlocality \cite{Navascues2007, Navascues2008}, where states are arbitrary and measurements are only limited by the distinction between the parties. However, when additional structure is present in the problem (constrains associated with e.g.~dimension, distance metrics,  energy, principal components etc.)~it is not straightforward  to efficiently incorporate it into NPA-type methods. This leaves an important knowledge gap: how can SDP relaxations efficiently approximate quantum correlations in the breadth of physics problems that involve more general structural constraints on the involved quantum resources?

In this work, we develop an overarching methodology to address this question. We propose an SDP relaxation methodology for quantum correlations which is compatible with a variety of relevant constraints on the involved quantum resources and is  adaptable to different physical scenarios. It reliably provides useful or even optimal correlation bounds at reasonable computational cost. By applying the methodology to address five different open research problems in quantum information science, we exemplify its versatility, its performance benchmarks and some of the physical scenarios and types of constraints for which it is useful. Specifically, we develop techniques to address the following five problems: 

\begin{list}{Problem \arabic{enumi}:}{
  \usecounter{enumi}
  \setlength{\leftmargin}{0.9cm}
}
	\item correcting entanglement witnesses for imperfect control of local measurement devices.
	\item certifying quantum measurement devices using states created from sources with bounded fidelity.
	\item bounding the dimensionality of genuine multipartite entangled states.  
	\item computing the dimension needed to operate a quantum state preparation device. 
	\item determining uncertainty relations for anticommuting observables with calibration errors. 
\end{list}
Table~\ref{Tab_overview_problems} provides a simple overview: we  list the physical scenario for each problem together with the characteristic constraints that it requires us to introduce into the  SDP relaxation. 
\begin{table*}[t]
	\centering
	\renewcommand{\arraystretch}{1.4}
	\begin{tabular}{|c|c|c|c|}
		\hline
		\textbf{Prob.} & \textbf{Scenario} & \textbf{Main constraints} & \textbf{Contribution} \\
		\hline
		\hyperref[sec:ent_witness]{1} & Bipartite entanglement witness & Dimension, separability \&  fidelity & Making entanglement witnesses robust to  errors \\
		\hline
		\hyperref[sec:distrust]{2} & Prepare-and-measure  & Fidelity in subspace & Characterising correlations from imperfect sources \\
		\hline
			\hyperref[sec:GME]{3}  & Multipartite entanglement source & Schmidt-rank \& fixed state & Enhanced multiparticle entanglement detection\\ 
		\hline
			\hyperref[sec:abs_dim]{4}  & Quantum preparation device & Subspace dimension \& fixed state & Witnessing the dimension of quantum states\\
		\hline
			\hyperref[sec:uncertainty]{5}  & Uncertainty relations & Operator norm & Correcting uncertainty relations for calibration errors\\
		\hline
	\end{tabular}
	\caption{The SDP relaxation methodology is used to address five different quantum information problems. The  scenario for each problem is listed together with the main types of constraints required to address it and the corresponding  contribution to the state-of-the-art. }\label{Tab_overview_problems}
\end{table*}

This article is structured as follows. In section~\ref{sec:methods} we present the overarching methodology for the SDP relaxations. In sections~\ref{sec:ent_witness}-\ref{sec:uncertainty} we apply the method to each of the problems 1-5 respectively. In section~\ref{sec:conclusions} we discuss the different features of the methodology highlighted by the results, as well as its broader relevance and its outlook as a tool in quantum information science.

\section{Methodology}\label{sec:methods}

The characterisation of quantum correlations can be viewed as a non-commutative polynomial decision problem over an appropriately constrained operator space. Consider a set of bounded operators $\{X_i\}_i\subset \mathcal{B}(\mathcal{H})$ on a Hilbert space $\mathcal{H}$. This can represent  e.g.~a probability distribution over the outcome statistics of an experiment, a collection of quantum states or some other type of quantum correlations. Our main question is to decide whether $\{X_i\}_i$ can be generated  with a given set of quantum resources. 

Formally, consider a set of bounded operators $L=\{O_1, \dots, O_n\}\subset \mathcal{B}(\mathcal{H}')$ acting on a Hilbert space $\mathcal{H}'$. These operators may represent the relevant quantum states, measurements or transformations used to generate the correlations $\{X_i\}_i$. Furthermore, let $S$ be a set whose elements are monomials over the operators in $L$ and let $\Lambda_i: \mathcal{B}(\mathcal{H}')\rightarrow \mathcal{B}(\mathcal{H})$ and $T_{j},R_{k}$ be linear maps. We define the decision problem
\begin{equation}\label{eq:problem_general}
	\begin{array}{ccc}
		\text{find}  & \qquad L=\{O_1,\ldots,O_n\} \\ [2ex]
		\text{s.t} & \ \Lambda_i \big(S(L)\big)= X_i & \forall i \\[0.5ex]
		& \ T_j \big(S(L)\big)\succeq 0 & \forall j \\[0.5ex]
		& \ R_k \big(S(L)\big)= 0 & \ \forall k. \\[0.5ex]
	\end{array}
\end{equation}
Here, $\Lambda_i$ are the linear maps that transform the monomials $S(L)$ into the correlations $X_i$. The linear maps $T_j$ and $R_k$ are respectively used to describe semidefinite and equality constraints that may apply to the monomials. In particular, polynomial optimisation in non-commutative variables can be viewed as as a special case of the above.

To prove an affirmative answer to the decision problem, one needs only to find a feasible set $L$, but proving a negative answer is often more challenging. Our goal is to address the latter by developing a sequence SDP relaxations of the set of feasible $L$. The sequence is a hierarchy, in the sense that every next step is at least as accurate as the former for providing an outer approximation of the quantum correlations. To this end, let us choose  $S(L)$ to consists of all products of length $0,1,\ldots, K$ over the operators in $L$. We refer to $K$ as the order of the monomials, adopting the convention that the zeroth order corresponds to the identity element. For instance, selecting $K=1$ corresponds to $S=\{L\}\cup\{\mathds{1}\}$ and selecting $K=2$ corresponds to $S=\{L\}\cup\{\mathds{1}, O_iO_j\}$, where it is for simplicity left implicit that the whole set $\{O_iO_j\}_{i,j}$ is included in $S$.  

To relax the original problem into an SDP, we introduce a completely positive map $\Theta$ and define a matrix of the form 
\begin{equation}\label{eq:block_mat}
	\Gamma=\sum_{u,v\in S}  \ketbra{i_u}{i_v}\otimes \Theta\left(u v^\dagger\right).
\end{equation} 
In the above, $i_u$ ($i_v$) corresponds to the position index of the monomial $u$ ($v$) in the list $S$ and $\Theta\left(u v^\dagger\right)$ represents variables of the relaxation.  The matrix $\Gamma$ can be seen as a block-matrix, where $\Gamma_{u,v}\equiv \Theta\left(u v^\dagger\right)$ is the block located at block-row $i_u$ and block-column $i_v$. Since these block-rows and block-columns are indexed by the monomials, we refer to $\Gamma$ as a \textit{block moment matrix} (BMM). The fundamental property of the BMM, which makes it ammenable to SDP methods, is that it is positive semidefinite by construction. This follows from 
\begin{equation}
	\begin{split}
		\Gamma&=\left(\mathds{1}\otimes \Theta\right) \left(\sum_{u}\ket{i_u}\otimes u\right)\left(\sum_{v} \bra{i_v}\otimes v^\dagger\right)\\
		&=\left(\mathds{1}\otimes \Theta\right) \left(MM^\dagger\right)\succeq 0,
	\end{split}
\end{equation}
where we have defined $M=\sum_{u}\ket{i_u}\otimes u$ and used that $\Theta$ is completely positive. Note that by choosing $\Theta(\cdot)=\bra{\psi}\cdot\ket{\psi}$ one recovers the NPA-type moment matrix.

In order for the BMM to be useful for relaxing the decision problem \eqref{eq:problem_general}, we must be able to recover the correlations $X_i$ by applying linear maps to it.  This means that we must choose  $\Theta$ such that for every $\Lambda_i$, there exists another linear map $\Omega_{\Lambda_i}$ with the property that $\Lambda_i=\Omega_{\Lambda_i}\circ \Theta$. Furthermore, $\Theta$ must be chosen in such a way that the constraints of the decision problem can be built into the BMM. It is helpful to think of the constraints as being of two different types.

The first type of constraints can be viewed as reduction rules: they originate from the algebraic structure of quantum theory and the properties of $\Theta$. These can imply either identities between some of the blocks  or algebraic restrictions on them. For example, if $u=Z$ and $v=Z'$, where $\{Z,Z'\}$ is a projective dichotomic measurement, orthogonality and projectivity respectively imply the reduction rules  $\Gamma_{Z,Z'}= 0$, $\Gamma_{Z,Z}= \Gamma_{Z,\mathds{1}}$ and $\Gamma_{Z',Z'}= \Gamma_{Z',\mathds{1}}$. Another example is if $\Theta$ is selected as the trace map. From the cyclicity of trace it follows that $\Gamma_{A,BC}=\Gamma_{B,CA}=\Gamma_{C,AB}$.

The second type of constraints are those associated with the maps $T_j$ and $R_k$ appearing in the decision problem \eqref{eq:problem_general}. In order to incorporate them in the BMM, we must find appropriate linear maps $\{\Omega_{T_j}\}_j$ and $\{\Omega_{R_k}\}_k$ acting on $\Gamma$ such that  $\Omega_{T_j}(\{\Gamma_{u,v}\}_{u,v})\succeq 0$ and $\Omega_{R_k}(\{\Gamma_{u,v}\}_{u,v})=0$ correspond to the original constraints.  This is most straightforward when one can choose $\Theta$ such that its action on the monomial products commutes with $T_j$ and $R_k$, i.e. $\Theta(T_j(S))= T_j(\Theta(S))$ and analogously for $R_k$. Under this property, one may formulate the constraints as relaxations in the image space of $\Theta$. However, some constraints could become trivial in the reduced space. To avoid this, we would have to require the stronger conditions $T_j=\Omega_{T_j}\circ\Theta$ and $R_k=\Omega_{R_k}\circ\Theta$. In practice, whether this strengthening is necessary is problem-dependent. 

We can now consider the BMM SDP relaxation of problem \eqref{eq:problem_general}. It reads 
\begin{equation}\label{eq:BMM_general_abstract}
	\begin{array}{ccc}
		\text{find} & \Gamma & \\[2ex]
		\text{s.t} & \ \Omega_{\Lambda_i} \big(\{\Gamma_{u,v}\}_{u,v}\big)= X_i & \forall i \\[0.5ex]
		& \ \ \Omega_{T_j} \big(\{\Gamma_{u,v}\}_{u,v}\big) \succeq 0 & \forall j \\[0.5ex]
		& \ \Omega_{R_k} \big(\{\Gamma_{u,v}\}_{u,v}\big)= 0 & \ \forall k, \\[0.5ex]
		& \Gamma \succeq 0.		
	\end{array}
\end{equation}
where we left implicit the constraints implied by the reduction rules. The positivity of $\Gamma$ ensures that every feasible solution of the original problem maps to a feasible point of the SDP. Naturally, however, whether an SDP formulation of the type \eqref{eq:BMM_general_abstract} is possible depends on whether one can find the appropriate choice of $\Theta$ for the specific problem under consideration. To gain further insight on this matter, we distinguish between two classes of problems, namely (i) those with a fixed Hilbert space dimension, and (ii) those without a fixed  Hilbert space dimension. 

For problems of type (i), we can always choose $\Theta=\text{id}$, where $\text{id}$ is the identity map, because the size of the blocks $\Gamma_{u,v}$ is given. We can then directly apply the linear maps $\Lambda_i, T_j, R_k$ to the relevant blocks of $\Gamma$ and thereby obtain an SDP relaxation to every decision problem \eqref{eq:problem_general}. This means that in Eq~\eqref{eq:BMM_general_abstract} the maps $\Omega_{\Lambda_i} $, $\Omega_{T_j} $ and $\Omega_{R_k}$ become just the maps  $\Lambda_i$ , $T_j$ and $R_k$ respectively, when applied  to the appropriate block. 

For problems of type (ii), $\Theta$ is used to compress the potentially infinite-dimensional Hilbert space $\mathcal{H}'$ into a finite-dimensional space in which the blocks in the BMM reside. As noted above, this limits which constraints can be imposed directly on the BMM and typically weakens the relaxation.  However, in the important case where the constraints are  polynomials in non-commutative variables,  they can be introduced into the SDP relaxation via  localising matrices.  For instance, to enforce $g(L) \succeq 0$, with $g$ representing a Hermitian polynomial, we can associate a localising BMM of the form $\bar{\Gamma}_{g}=\sum_{u,v}\ketbra{i_u}{i_v}\otimes \Theta\left(u^\dagger g v\right)$. The localising matrix is positive semidefinite by construction if the polynomial is positive. By choosing appropriately the order of the monomials, $K_g$, used to build $\bar{\Gamma}$, the localising matrix can be expressed as a linear map over $\Gamma$. Hence, it yields additional constraints without introducing additional variables.

\subsection{Adding an objective function}
Our discussion for the decision problem in Eq~\eqref{eq:problem_general} and its corresponding SDP relaxation can straightforwardly be modified to apply also to situations where we have an objective function instead of a fixed set of correlations $\{X_i\}$. This is a natural setting frequently encountered in quantum correlation studies. Consider an objective function of the  general form
\begin{equation}\label{object}
W\left(\{X_i\}\right)=\sum_{i} \tr\left(F_iX_i\right),
\end{equation}
for some Hermitian $F_i$. Thus, instead of reproducing the correlations $\{X_i\}$, we seek to maximise $W$ over the space of feasible $\{X_i\}_i$. Specifically, we want to determine upper bounds on the solution to
\begin{equation}\label{eq:problem_general_objective}
\begin{array}{ccc}
\text{max}  & \qquad W\left(\{X_i\}\right) \\ [2ex]
\text{s.t} & \ \Lambda_i \big(S(L)\big)= X_i & \forall i \\[0.5ex]
& \ T_j \big(S(L)\big)\succeq 0 & \forall j \\[0.5ex]
& \ R_k \big(S(L)\big)= 0 & \ \forall k, \\[0.5ex]
\end{array}
\end{equation}
where ${X_i}$ are free variables. Since each $X_i$ can be obtained by linear operations on the BMM, the same also applies to $W\left(\{X_i\}\right)$. Thus, it can be introduced into the SDP relaxation in place of specifying each $X_i$.

In the rest of this work, we apply the general methodology presented above to concrete, physically motivated, open problems in quantum information  and show how it can be used to solve them.
However, before doing so, we first give a simple warm-up example illustrating how to use the methodology.

\subsection{An illustrative example}
We consider a simple example to illustrate how the BMM SDP relaxation can  directly account for constraints at the operator level.  Consider the maximisation problem
\begin{equation}\label{eq:problem_pauli}
\begin{array}{cc}
\max_{O_1,O_2} & \tr(O_1+O_2) \\[2ex]
\text{s.t.} & O_1^2=\mathds{1}, \quad O_2^2=\mathds{1}, \\[0.5ex]
& [O_1, O_2]=2i\begin{bmatrix}
1 & 0\\ 0 & -1
\end{bmatrix}\\[0.5ex]
&O_1,O_2\in\text{Herm}_2,
\end{array}
\end{equation}
defined over the Hermitian $2\times 2$ matrices. We may recognise the constraints as describing Pauli observables $O_1=\sigma_X$ and $O_2=\sigma_Y$, leading to  $\tr\left(O_1+O_2\right)=0$. Nevertheless, let us suppose we do not know this and instead want to solve the problem through SDP relaxation using the BMM method. For that, consider the first-order ($K=1$) monomial list $S=\{\mathds{1}, O_1, O_2\}$ and choose $\Theta=\text{id}$. From Eq~\eqref{eq:block_mat} the BMM is 

\begin{equation}
\Gamma=\begin{pmatrix}
\mathds{1} & O_1 & O_2 \\
& \mathds{1} & R \\
& & \mathds{1} 
\end{pmatrix},
\end{equation}
where we show only the upper diagonal because $\Gamma$ is Hermitian. Here, we have used the constraint $O_i^2=\mathds{1}$ from Eq~\eqref{eq:problem_pauli} when considering the blocks $\Gamma_{O_1,O_1}$ and  $\Gamma_{O_2,O_2}$. For the blocks $\Gamma_{\mathds{1},O_1}=O_1$ and $\Gamma_{\mathds{1},O_2}=O_2$, we assign Hermitian variables. For the block $\Gamma_{O_1,O_2}=O_1O_2^\dagger$, we assign the matrix variable $R$.  The commutator constraint in Eq~\eqref{eq:problem_pauli} can now be expressed as a linear equality over the BMM, i.e~as $R-R^\dagger=2i\sigma_Z$. Under these constraints and with $\Gamma\succeq 0$, we can evaluate as an SDP an upper bound on the objective function $\tr\left(O_1+O_2\right)$. The result is zero, as expected.

\section{Entanglement detection with imperfect measurements}\label{sec:ent_witness}

Detecting the entanglement of an initially unknown quantum state $\rho_{AB}$ is an essential capability for quantum technology. The most common method is entanglement witnessing  \cite{G_hne_2009}. In an entanglement witness test, Alice and Bob are instructed to perform specific local measurements on their respective shares of $\rho_{AB}$ and check its entanglement by evaluating a suitable witness parameter whose value is limited for separable states. The validity of such a test hinges on the assumption that Alice and Bob precisely perform the instructed measurements. However, this is an idealisation to which experiments can only aspire \cite{Cao2024}. A series of works have shown that already small alignment errors in the measurements can undermine the conclusions of well-known  entanglement witnesses and lead to false positives \cite{Rosset2012, MorelliEntImprecise, Tavakoli2024}. At the same time, analytical methods to correct entanglement witnesses for bounded measurement imperfections  are limited to convenient special cases and  available numerical methods significantly overestimate the required correction even for the simplest cases \cite{MorelliEntImprecise}. Below, we first formalise this problem and then we show how it can be addressed using BMM SDP relaxations.

\subsection{Description of Problem 1}
An entanglement witness is an inequality of the form
\begin{equation}\label{entwitness}
W=\sum_{a,b,x,y} c_{abxy} \tr(A_{a|x}\otimes B_{b|y}\rho_{AB}) \leq \beta,
\end{equation}
where $c_{abxy}$ are real coefficients, $\beta$ is the largest value of $W$ achievable with separable states, and $\{A_{a|x}\}$ and $\{B_{b|y}\}$ are the measurements of Alice and Bob where $x$ and $y$ label the measurement choices and $a$ and $b$ label the outcomes. A violation of the inequality implies that $\rho_{AB}$ is entangled. Importantly, Alice and Bob are given specific instructions to choose their measurements as the bases $\tilde{A}_{a|x}$ and $\tilde{B}_{b|y}$, so that the inequality \eqref{entwitness} is valid. However, consider that their actual lab realisations of these target  measurements are not perfect, but only close approximations \cite{Rosset2012}. The quality can be benchmarked by the fidelity bounds \cite{MorelliEntImprecise}
\begin{align}\label{eq:meas_dist}\nonumber
&\tr\big(A_{a|x}\tilde{A}_{a|x}\big) \geq 1-\varepsilon^A_{a|x}\\
& \tr\big(B_{b|y}\tilde{B}_{b|y}\big) \geq 1-\varepsilon^B_{b|y},
\end{align}
where $0\leq \varepsilon^A_{a|x},\varepsilon^B_{b|y}\ll 1$ represent the imperfection for each measurement choice and outcome. Let us label the set of imperfection parameters as $\vec{\varepsilon}=\{\varepsilon^A_{a|x},\varepsilon^B_{b|y}\}_{a,b,x,y}$. 
Setting  $\vec{\varepsilon}=0$ corresponds to $A_{a|x}=\tilde{A}_{a|x}$ and $B_{b|y}=\tilde{B}_{b|y}$; thus the standard entanglement witness in Eq~\eqref{entwitness} can be used. For non-ideal measurements, for which $\vec{\varepsilon}\neq 0$, the bound $\beta$ must be updated into an $\vec{\varepsilon}$-dependent bound $\beta(\vec{\varepsilon})$, thereby eliminating potential false positives. The task is to address $\beta(\vec{\varepsilon})$ by bounding from above the left-hand-side of Eq~\eqref{entwitness} for all separable states and all projective measurements satisfying the conditions \eqref{eq:meas_dist} in the given Hilbert space.

\subsection{BMM method for solving Problem 1}
We now show how the BMM SDP relaxations solve the problem. Because the degrees of freedom are known, the problem is the defined over a bipartite tensor-product Hilbert space $\mathcal{H}'=\mathcal{H}_{A}\otimes\mathcal{H}_{B}$ where  $d=\text{dim}(\mathcal{H}_A)=\text{dim}(\mathcal{H}_B)$ is given. Therefore, we consider a BMM whose block-dimension is $d^2$ (equal to the global Hilbert space dimension), corresponding to using the identity map $\Theta=\text{id}$ in Eq~\eqref{eq:block_mat}. To lighten the notation, in what follows we omit tensor product with the identity when extending local operators (e.g. we write $\rho_A$ instead of $\rho_A\otimes\mathds{1}$). Furthermore, since we are interested in separable states and the witness is linear, it is sufficient to consider product states. This means $\rho_{AB}=\rho_A\rho_B$, where $[\rho_A,\rho_B]=0$. We identify the operators relevant to the problem as $L=\{\rho_A, \rho_B,A_{a|x}, B_{b|y}\}$ and use $L$ to construct the BMM. 

To illustrate  the reasoning behind the construction of the BMM, we choose the monomial list $S=\left\lbrace \mathds{1}, \rho_A, \rho_B, \rho_A\rho_B, A_{a|x}B_{b|y} \right\rbrace$ as an example. The BMM takes the block-form
\begin{equation}\label{eq:BMM_witness}
	\Gamma=\begin{pmatrix}
		\mathds{1} & \rho_A & \rho_B & \rho_A\rho_B & \{A_{a|x}B_{b|y}\}  \\
		& Q_1 & Q_2 & Q_3 & P_1 \\
		& & Q_4 & Q_5 & P_2 \\
		& & & Q_6 & P_3 \\
		& & & & R
	\end{pmatrix},
\end{equation}
where we denoted by $Q_i$ all $d^2\times d^2$ complex matrix variables that correspond to products of states; by $P_i$ the row-vectors of matrix variables that represent products of states and measurements, with concrete elements $P_1^{abxy}=\rho_A A_{a|x}B_{b|y}$, $P_2^{abxy}=\rho_B A_{a|x}B_{b|y}$ and $P_3^{abxy}=\rho_{A}\rho_{B} A_{a|x}B_{b|y}$; and by $R$ the array containing blocks that represent only products of measurement operators, with elements $R^{abxy}_{a'b'x'y'}=A_{a|x}B_{b|y}A_{a'|x'}B_{b'|y'}$.

Let us now see how the relevant constraints of the problem are introduced to the BMM. First, since our goal is to maximise the value of $W$ in Eq~\eqref{entwitness}, we can restrict $\rho_A$ and $\rho_B$ to pure states, i.e.~$\rho_A^2=\rho_A$ and $\rho_B^2=\rho_B$. This together with the product structure implies the reduction rules  $Q_{2}=Q_{3}=Q_{5}=Q_{6}=\rho_A\rho_B$, $Q_{1}=\rho_A$ and $Q_{4}=\rho_B$. Also, we impose the constraint $\tr(Q_1)=\tr(Q_4)=1$ since states have unit trace.  Next, we need to impose the separability of the shared state. This is already implicit in our factorisation of $\rho_{AB}$ but it has further implications on the BMM. Since separability admits no semidefinite characterisation \cite{fawzi2019}, we instead consider that the state remains positive under  partial transposition  \cite{PPT}. This means  $Q_2^{T_A}\succeq 0$.   Consider now the array $R$. Because $A_{a|x}$ and $B_{b|y}$ are commuting projective measurements, it simplifies as $R_{a'b'xy}^{abxy}=\delta_{a,a'}\delta_{b,b'}A_{a|x}B_{b|y}$. We can then consider the linear combination of blocks $R_{a\lvert x} =\sum_{a',b',b} R_{a'b'xy}^{abxy}=A_{a\lvert x}$ and constrain it to be positive semidefinite. It also lets us  impose  the fidelity constraint  \eqref{eq:meas_dist} as $\tr\big(R_{a\lvert x} \tilde{A}_{a|x}\big)\geq d(1-\varepsilon^A_{a\lvert x})$. The analogous applies to Bob's measurements. 
Finally, we can construct the objective function, namely the witness in Eq~\eqref{entwitness}, as a linear functional over the block $P_3$ in the BMM. Specifically, we have  $W=\sum_{a,b,x,y}c_{abxy}\tr(P_3^{abxy})$.  In summary, the SDP relaxation corresponds to maximising this expression of $W$ while imposing the above constraints such that the BMM in Eq~\eqref{eq:BMM_witness} is positive semidefinite. The solution yields an upper bound on $\beta(\vec{\varepsilon})$.

\subsection{Results for Problem 1}
We now showcase the performance of the SDP relaxation method. To this end, consider the well-known entanglement witness test in which  Alice and Bob perform global product measurements in two mutually unbiased bases \cite{Spengler2012}. A typical choice of these bases is  the computational basis and the Fourier basis, i.e.~the target measurements are $\tilde{A}_{k|0}=\tilde{B}_{k|0}=\ketbra{k}{k}$ and  $\tilde{A}_{k|1}=\tilde{B}_{k|1}=\ketbra{f_k}{f_k}$, where $\ket{f_k}=\frac{1}{\sqrt{d}}\sum_{j=0}^{d-1}\omega^{kj}\ket{j}$ with $\omega=e^{\frac{2\pi i}{d}}$. The entanglement witness then takes the  form \cite{Spengler2012}
\begin{align}\label{eq:Pauli_wit2}
	W = \sum_{j=0,1}\sum_{k=0}^{d-1}\tr(A_{k|j}\otimes B_{k|j}\rho)\leq 1+\frac{1}{d}.
\end{align}
We will update the bound for when the measurements no longer perfectly correspond to the targets. 

\begin{figure}
	\centering
	\includegraphics[width=1\linewidth]{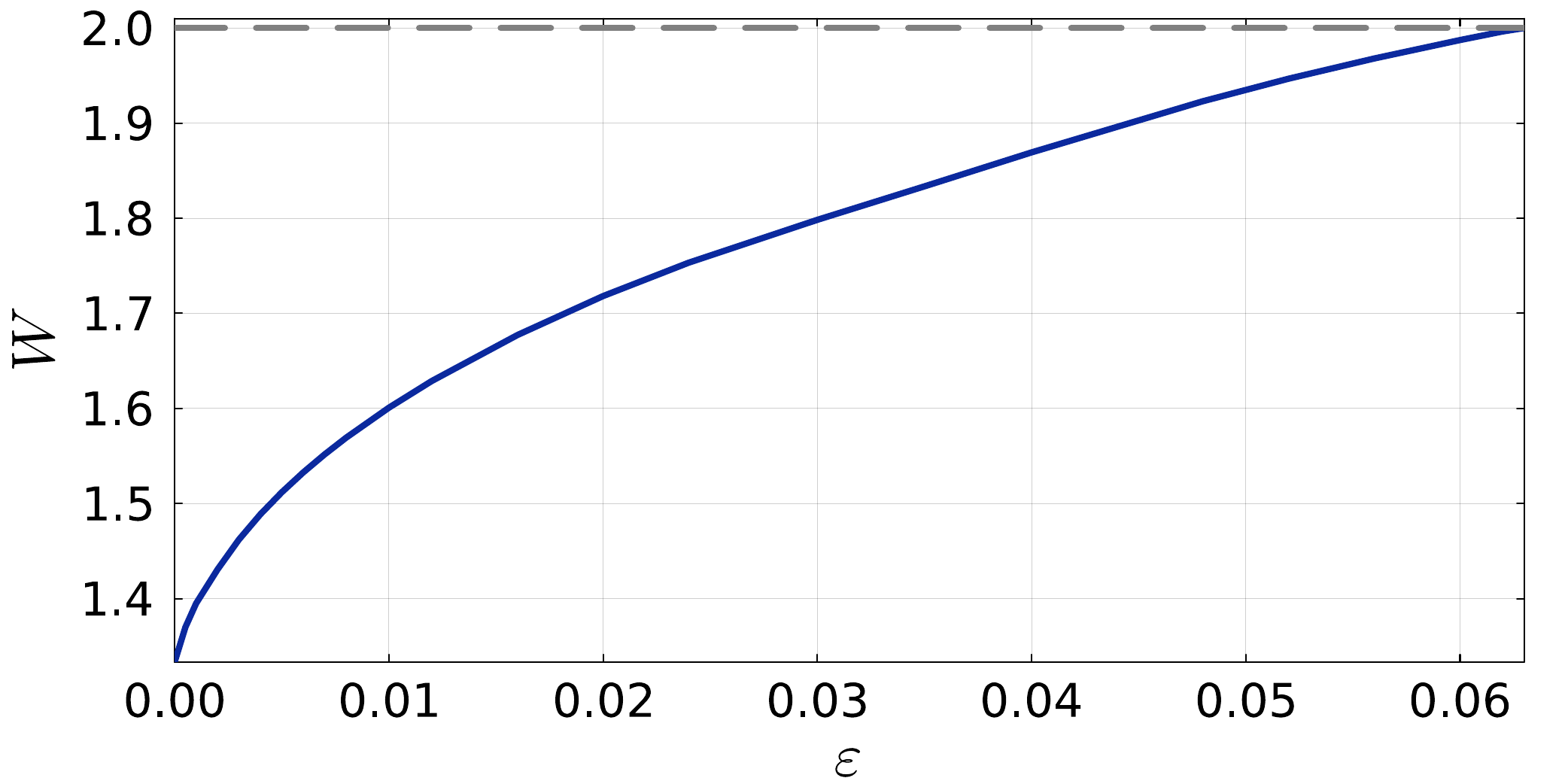}
	\caption{Upper bound on the entanglement witness in Eq~\eqref{eq:Pauli_wit2} for three-dimensional systems and imperfection parameter $\varepsilon$. The dashed line represent the maximal quantum value $W=2$. At $\varepsilon=0$ the witness bound is  $W=1+1/3$.}
	\label{Fig:ent_imprec_pauli}
\end{figure}

In our implementation, we have used the monomial list $S=\{L\}\cup \{\mathds{1}, \rho_A\rho_B, \rho_A A_{a|x}, \rho_B B_{b|y} \}$ which is intermediate between the first and second level ($K=1$ and $K=2$).  As a first check, we set all imperfection parameters equal, namely $\varepsilon^A_{a|x}=\varepsilon^B_{b|y}\equiv \varepsilon$, and considered qubit systems ($d=2$). Evaluating the SDP gives an updated bound in \eqref{eq:Pauli_wit2} identical to that  derived  analytically in Ref~\cite{MorelliEntImprecise} for any $\varepsilon$. This shows the method's potential for accuracy, but does not provide new knowledge. Secondly, we go further and show that the method also obtains new and useful results. For that, we consider experimentally more realistic situations \cite{Cao2024}, where the estimated values of $\vec{\varepsilon}$ are not uniform but instead measurement-dependent.  For such general $\vec{\varepsilon}$, theoretical solutions are not known in the literature, but from a series of random case studies we find that our method provides tight bounds on $\beta(\vec{\varepsilon})$. Note that this requires no increased computational cost.  Thirdly, we have used the same choice of $S$ to consider the case of qutrits ($d=3$), for which no analytical or numerical solution is known even when $\varepsilon^A_{a|x}=\varepsilon^B_{b|y}\equiv \varepsilon$. Our results are illustrated in Fig~\ref{Fig:ent_imprec_pauli}. For small $\varepsilon$, which is the practically most relevant regime, the bound provides a sizable gap with the value $W=2$  achievable by entangled states. Finally, another useful feature of the SDP approach is that it does not have to deal with a specific witness, but can apply directly to any experimentally measured probability distribution. This amounts only to imposing the experimental probabilities  as additional linear constraints in the SDP; in the example based on \eqref{eq:BMM_witness} this amounts to constraining the trace of the blocks in the array $P_3$. It allows for more powerful entanglement detection from the same data.  Notice that while we should in general consider mixed states when considering compatibility with experimental probabilities, the convexity of the set of feasible $\Gamma$ allows us to remain with considering only pure states on the level of the SDP relaxation. 

\section{Quantum certification in the prepare-and-measure scenario}\label{sec:distrust} 
Certifying the functionality of quantum devices is a basic task for quantum technology. Here, we consider certification of quantum measurement devices \cite{Guhne2023}. This is straightforward via detector tomography, but that requires access to a flawless source of states for probing the device. The practical issues associated with the latter has motivated semi-device-independent proposals where  certification is based only on the system's dimension \cite{Tavakoli2018, Farkas2019, Carmeli2020, Navascues2023}. However, this has two significant drawbacks:  assuming no control in the given dimensions is excessively pessimistic for state-of-the-art quantum control, and it is rarely the case that a system can be flawlessly described in a finite-dimensional Hilbert space  (see e.g.~\cite{jef_almost_qudit}).  To overcome both of these issues, one can instead consider certifying measurement devices based on more well-adapted assumptions. For instance, one may view the source of potentially infinite-dimensional, but assume that the fidelity it achieves with the state it was designed to generate is high and bounded from below.  While this is operationally more natural, characterising the correlations is harder. Available methods are limited to the few simplest scenarios, rely on sampling heuristics and make use of  additional unwarranted  assumptions \cite{DistrustTavakoli}. Here, we show how quantum certification with fidelity-bounded sources can be achieved using the BMM SDP relaxations that overcome these three drawbacks.

\subsection{Description of Problem 2}
Consider a prepare-and-measure scenario where the sender, Alice, selects an input $x$ and prepares a probe state $\rho_x$ that is sent to the receiver, Bob. Bob selects an input $y$ and performs an associated measurement $\{M_{b|y}\}_b$, where $b$ is the outcome. The measurement device is treated as uncharacterised. Moreover,  we assume that Alice and Bob's devices can share classical correlations. This is represented by a random variable $\lambda$ with some distribution $q(\lambda)$. Therefore, the correlations observed in the experiment are given by 
\begin{equation}\label{eq:PM_corr}
	p(b|x,y)=\sum_\lambda q(\lambda)\tr\left(\rho_x^{(\lambda)}M^{(\lambda)}_{b|y}\right) .
\end{equation}
Alice's goal is to prepare some well-selected target state $\ket{\psi_x}$. However, she cannot perfectly control her source to produce it. The actually produced state, $\rho_x=\sum_\lambda q(\lambda) \rho_x^{(\lambda)}$, can live in a larger Hilbert space and is only be assumed to have a high fidelity with $\ket{\psi_x}$.  Thus, Alice's source satisfies \cite{DistrustTavakoli} 
\begin{equation}\label{eq:epsilon_fid}
\bracket{\psi_x}{\rho_x}{\psi_x}\geq 1-\omega_x \ ,
\end{equation}
where $\ket{\psi_x}$ should be seen as embedded into the larger Hilbert space of $\rho_x$ and  $\omega_x$ is the distrust parameter representing Alice's lack of perfect control for the $x$'th probe state. The task is to bound the  space of correlations $p(b|x,y)$ achievable under given distrust parameters $\{\omega_x\}_x$, both when considering general quantum measurements and when limiting the measurements to those that admit a classical representation. In the latter case, we must constrain the measurements further, so that there are  orthonormal bases $\{|e_k^{(\lambda)}\rangle\}_{k,\lambda}$ from which the measurements can be obtained via a classical post-processing, 
\begin{equation}\label{clmeasurement}
M^{(\lambda)}_{b|y}=\sum_{k} p(b|y,k,\lambda)\ketbra*{e_k^{(\lambda)}}{e_k^{(\lambda)}}
\end{equation}
for some probability distribution $p(b|y,k,\lambda)$. This can equivalently be seen as Bob's measurements commuting when conditioned on $\lambda$.

\subsection{BMM method for solving Problem 2}
A key feature of this problem is that the relevant Hilbert space $\mathcal{H}'$ of Alice's states and Bob's  measurements is potentially  unbounded, but that the fidelity constraint \eqref{eq:epsilon_fid} applies with respect to the fixed $d$-dimensional subspace in which Alice's target states live. Therefore, the guiding intuition is build the BMM using, essentially, blocks of size  $d\times d$  and to choose the map $\Theta$ in such a way that it eliminates all other parts of the Hilbert space. To this end, we select  $\Theta=\text{id}_d\oplus\tr_{\perp}$, where $\perp$ is the part of $\mathcal{H}'$ complementary to the $d$-dimensional subspace of $\{\ket{\psi_x}\}_x$.  The fidelity condition \eqref{eq:epsilon_fid} is then  expressed as
\begin{equation}\label{blockfid}
	\bracket{\psi_x}{\Theta(\rho_x)}{\psi_x}\geq 1-\omega_x,
\end{equation}
which is a useful formulation because $\Theta(\rho_x)$ will appear as $(d+1)$-dimensional  blocks in the BMM.

To illustrate the BMM construction, we first identify the operators of Alice and Bob to be $L=\{\rho_x, M_{b|y} \}$. For simplicity of illustration, we consider the first relaxation level ($K=1$), corresponding to monomials $S=\{L\}\cup \{\mathds{1}\}$. The BMM takes the form 
\begin{equation}\label{eq:BMM_PAM}
	\Gamma=\begin{pmatrix}
		\mathds{1}_d\oplus \text{d}_\perp & \{\Theta(\rho_x)\} & \{\Theta(M_{b|y})\}  \\
		& Q & P \\
		& & R 
	\end{pmatrix} ,
\end{equation}
where $Q$, $P$ and $R$ are arrays  composed of the blocks $Q_{x'}^{x}=\Theta(\rho_{x'}\rho_{x})$, $P_{x}^{by}=\Theta(\rho_{x}M_{b|y})$, and $R_{b'y'}^{by}=\Theta(M_{b'|y'}M_{b|y})$ respectively, $\mathds{1}_d$ is the identity in the $d$-dimensional target space  and $\text{d}_\perp$ is a scalar variable representing the number of unknown dimensions that are traced out. Note that every block in the BMM is a direct sum of a $d\times d $ matrix and a  scalar.

Let us now survey the relevant constraints. Firstly, since the dimension is unbounded, measurements can be assumed projective without loss of generality. This means $M_{b|y}M_{b'|y}=\delta_{b,b'}M_{b|y}$ which implies $R_{by}^{b'y}=\delta_{b,b'}\Theta(M_{b|y})$. We can also impose normalisation as a trace constraint on the blocks because $1=\tr(\rho_x)=\tr\left(\Theta(\rho_x)\right)$ and $\sum_b\Theta(M_{b|y})=\mathds{1}_d\oplus\text{d}_\perp$.  Secondly, one can restrict Alice's states to be pure by imposing $\rho^2_x= \rho_x$ but this is in general not a legitimate restriction in prepare-and-measure scenarios  (see e.g.~\cite{Tavakoli2021Contextual,Chaturvedi2021characterising, Tavakoli2022informationally}). To allow for mixed states, the relevant constraint is $\rho_x \succeq \rho_x^2$, which can be introduced as the semidefinite  constraint $\Gamma_{\mathds{1},\rho_x}\succeq\Gamma_{\rho_x,\rho_x}$ on the BMM.  Thirdly, the fidelities in Eq~\eqref{blockfid} are  linear inner-product constraints over the blocks $\Gamma_{\mathds{1},\rho_x}$. Fourthly, if we want to restrict to classical measurements we must additionally introduce commutation to Bob's measurements. That implies  $R_{b'y'}^{by}=R_{by}^{b'y'}$. Finally, the probabilities \eqref{eq:PM_corr} can be expressed as $p(b|x,y)=\tr(P_{x}^{by})$. The feasibility of  $\Gamma\succeq 0$ under these constraints is a necessary condition for $p(b|x,y)$ to admit a quantum (or a classical, respectively) realisation for the given distrust parameters $\{\omega_x\}_x$. 

Lastly, let us comment on the role of the random variable $\lambda$. Imagine momentarily that we had defined the BMM \eqref{eq:BMM_PAM} and all the above constraints for a given $\lambda$ and labeled it $\Gamma_{\lambda}$. We could then define a new BMM as the convex combination $\Gamma= \sum_\lambda q(\lambda) \Gamma_{\lambda}$, which is positive semidefinite by construction. Since the constraints on $\Gamma_{\lambda}$ are linear, they carry over to $\Gamma$, thereby leading to the same SDP relaxation based on $\Gamma$. Therefore, shared classical randomness is indirectly accounted for in our SDP relaxation.

\begin{figure}
	\centering
	\includegraphics[width=1\linewidth]{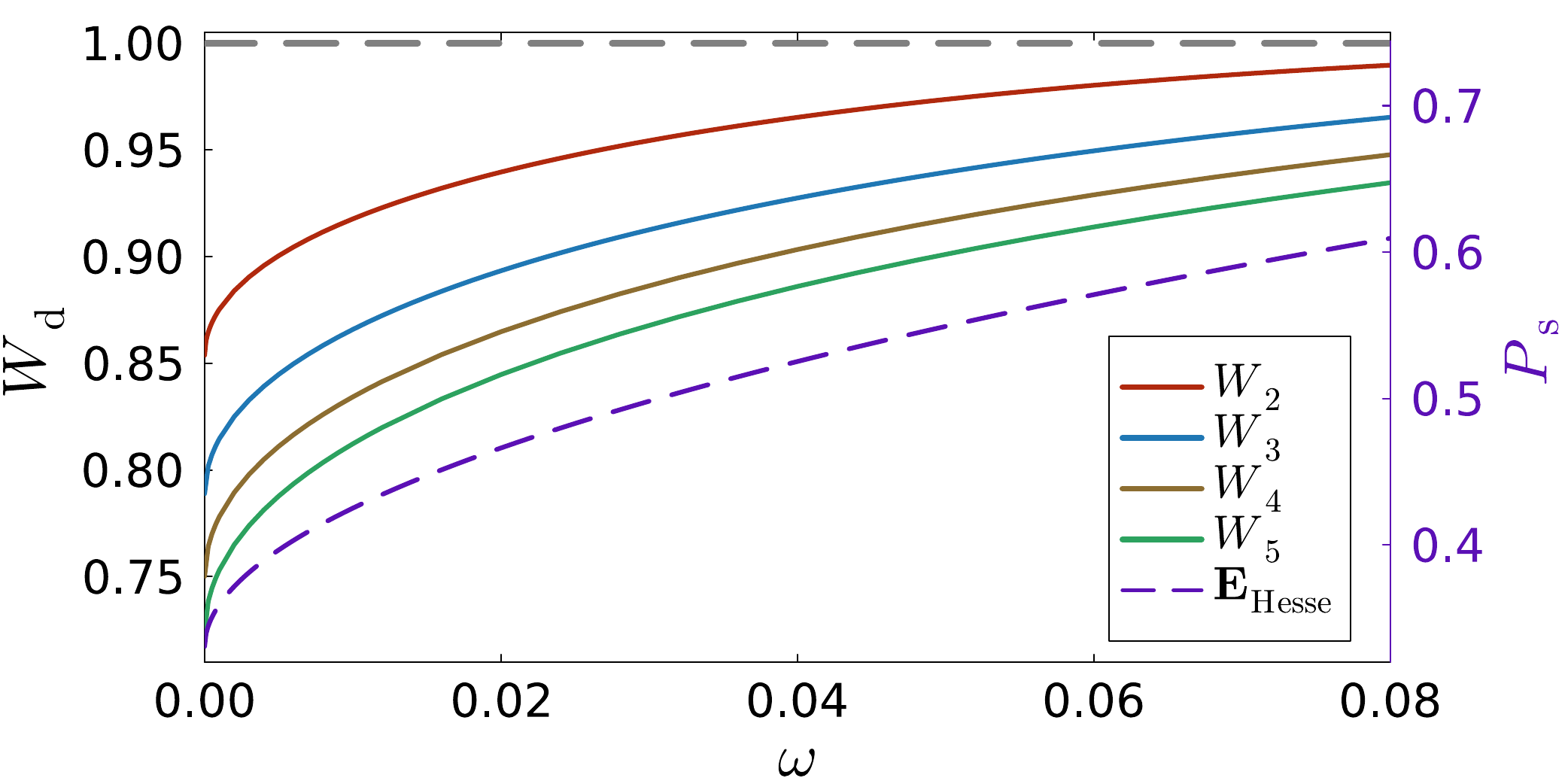}
	\caption{Upper bounds on the classical value of the witness $W_{d}$ for dimensions $d=2,\dots, 5$ (left axis) and success probability for discriminating the Hesse SIC $\textbf{E}_{\text{Hesse}}$ (right axis) as function of the distrust. No assumption is made on the communication channel's dimension. The dashed gray line marks the witness' quantum bound. }
	\label{Fig:plot_RAC_d5}
\end{figure}

\subsection{Results for Problem 2}

We benchmark the SDP relaxation by certifying quantum features in the measurement device. To this end, we consider the frequently employed pair of measurement bases corresponding to the $d$-dimensional computational basis and its Fourier transform, which are mutually unbiased.  A natural approach to certify that a device can implement these measurements is to probe each basis with states prepared in one of its $d$ eigenstates. To formalise this, we let Alice's input take the form $x=x_1x_2$ where $x_1\in\{0,1\}$ and $x_2\in \{0,1,\dots,d-1\}$. Her target states $\ket{\psi_{0x_2}}$ correspond to the computational basis $\ket{x_2}$ and the target states $\ket{\psi_{1x_2}}$ correspond to the Fourier basis $\ket{f_{x_2}}$. Bob has two inputs, $y\in\{0,1\}$ which he uses to read out the value of $x_2$ when $y=x_1$.  The relevant figure of merit becomes
\begin{equation}\label{eq:mixed_RAC}
	W_{d}=\dfrac{1}{2d}\sum_{x}p(b=x_2|x,y=x_1) \ .
\end{equation}
If Bob's device can implement both the intended measurements, it will measure the computational and Fourier bases and achieve a perfect score of $W_{d}=1$. However, if Bob's device admits a classical description \eqref{clmeasurement}, then  the possible values of $W_d$ are limited. For an ideal source ($\omega_x=0$) we have checked that for $d=2,3,4,5$ this limit is $W_{d}=\frac{1}{2}\left(1+\frac{1}{\sqrt{d}}\right)$. We now use the BMM method to evaluate this bound for non-ideal sources. A violation of it thereby certifies quantum features in the measurement device.

We  select the monomial list $S=\{\mathds{1}, \rho_x,M_{b\lvert y}, \rho_xM_{b|y} \}$ and implement the BMM SDP relaxation for $d=2,3,4,5$. The results are illustrated in \figref{Fig:plot_RAC_d5}. As we increase the dimension of $\{\ket{\psi_x}\}$, the separation between classical and quantum becomes larger. The quantum advantage (up to four decimals) persists up to $\omega \simeq 0.14$, $\omega\simeq0.20$, $\omega\simeq 0.24$ and $\omega\simeq 0.26$ for $d=2,3,4$ and $5$, respectively. Moreover, for small $\omega$, all the reported upper bounds are nearly optimal, with the gap increasing somewhat for larger $\omega$. For instance, the worst-case discrepancy over all $d$  at $\omega=0.01$ and at $\omega=0.05$ is no larger than $4.1\times 10^{-3}$ and $5.0\times 10^{-3}$. This was investigated using alternating convex search methods to obtain lower bounds on $W_{d}$. 

The above exemplifies results based on classical measurements. Let us also showcase the performance in a scenario with quantum measurements. For this, we consider a state discrimination task targeting a set of three-dimensional states that is symmetric and informationally complete \cite{Renes2004}. This means selecting $\{\ket{\psi_x}\}_{x=1}^{9}$ such that $\left\lvert\braket{\psi_x}{\psi_{x'}}\right\lvert^2=\frac{1}{4}$ for $x\neq x'$. The standard example of such a construction is the so-called Hesse SIC, whose (unnormalised) vectors are the columns of the matrix
\begin{equation}
\begin{pmatrix}
	0 & 0 & 0 & -1 & -\theta^{2} & -\theta & 1 & \theta & \theta^{2} \\
	1 & \theta & \theta^{2} & 0 & 0 & 0 & -1 & -\theta^{2} & -\theta \\
	-1 & -\theta^{2} & -\theta & 1 & \theta & \theta^{2} & 0 & 0 & 0
\end{pmatrix}
\end{equation}
with $\theta=e^{2\pi i/3}$. The success probability of discrimination becomes $P_\text{s}=\frac{1}{9}\sum_{x} p(b=x\lvert x)$, which for the target states is $P_\text{s}=\frac{1}{3}$. Using the monomials $S=\{\mathds{1}, \rho_x, \rho_xM_{x|y}, M_{x|y}\rho_x\}$ we have evaluated upper bounds on $P_\text{s}$ via the SDP method and the results are illustrated by the dashed line in Fig~\ref{Fig:plot_RAC_d5}. The results are optimal, matching up to numerical precision the explicit models obtained through alternating convex search methods. Furthermore, we have compared the bound with the analytical upper bound derived proposed in \cite{Pauwels2025} and found that ours is strictly stronger. This difference is also significant: for instance, if we compare the corrections to $P_s$ due to the presence of distrust at $\omega=0.01$, we find that the maximum success probability is $\sim13\%$ lower than the prediction of the previously best bound.

\section{Dimensionality of multipartite entanglement}\label{sec:GME}

While determining whether an arbitrary state is entangled or separable is already hard \cite{Gurvits2003, Gharibian2009}, a further challenge stems from determining the entanglement dimensionality of states with multiple local levels. For multipartite systems, the problem scales rapidly when increasing dimensionality or the number of subsystems. Multipartite entanglement dimensionality can be addressed by considering the standard bipartite entanglement dimension (Schmidt number) \cite{Terhal2000} accross all possible  bipartitions of the system \cite{Huber2013}; with the smallest Schmidt number needed to generate the multipartite state being called the genuine multipartite entanglement (GME) dimension \cite{Cobucci2024}. The main analytical criteria  for the GME-dimension are based on fidelity  with a target state, but this has two drawbacks: most states are not detected by fidelity criteria \cite{Weilenmann2020} and for general states it is rarely straightforward to find the best state with respect to which  one should evaluate the fidelity.  Numerical methods  can do better, but they become too expensive on standard computers already beyond for example four-qutrit systems \cite{Cobucci2024}. In parallel, experiments have demonstrated high-dimensional multipartite entanglement between three qutrits \cite{Erhard2018, Cervera2022} and recently between four qutrits \cite{Bao2023, Hu2025}. We show that the BMM SDP relaxations can be used as a tool to bound the GME-dimension for high-dimensional multipartite states when the system's size is in the regime of near-future experiments. The method remains efficient where previous numerical methods become too expensive and it systematically outperforms the standard fidelity criteria.

\subsection{Description of Problem 3}
Consider a pure bipartite state $\ket{\psi}_{AB}$. Its entanglement dimension $r(\psi)$ is given by the rank of the reduced state, namely $\text{rank}(\psi_A)$. This is extended to mixed states $\rho_{AB}$ by the Schmidt number, which is defined as \cite{Terhal2000} 
\begin{align}\nonumber
	r(\rho_{AB})\equiv \min_{\{q_i\},\{\psi_i\}} \Big\{&r_\text{max}: \quad \rho_{AB}=\sum_i q_i \psi_i\\
	&\text{and}\quad r_\text{max}=\max_i r(\psi_i)\Big\},
\end{align}
where $\psi_i$ are pure states and $q_i$ are probabilities. Hence, the Schmidt number is the largest $r$ found among the pure states appearing in the optimal decomposition of $\rho_{AB}$. 

Consider now that we instead have a multipartite state  $\rho_{1\ldots n}$ with $n$ subsystem. Let $(P\lvert \bar{P})$ be some bipartition of the subsystems. The state is said to be genuinely multipartite entangled if it cannot be decomposed into a mixture of states separable with respect to some bipartition, namely \cite{G_hne_2009}
\begin{equation}\label{gmeeq}
	\rho=\sum_{(P\lvert\bar{P})} q_{(P\lvert\bar{P})} \sigma_{(P\lvert\bar{P})},
\end{equation}
where $q_{(P\lvert\bar{P})}$ is a probability  and $\sigma_{(P\lvert\bar{P})}$ is a state separable with respect to $(P\lvert\bar{P})$. However, we are interested in GME states that also have a genuinely high dimension. This is addressed by the GME dimension \cite{Cobucci2024}. The GME dimension of $\rho_{1\ldots n}$ is the smallest integer $r$ such that a decomposition of the form \eqref{gmeeq} is possible where each $\sigma_{(P\lvert \bar{P})}$ is a state with  Schmidt number at most $r$. Note that the special case of $r=1$ reduces the GME dimension to standard GME commonly used for many-qubit systems.

For pure states it is straightforward to determine the GME dimension, but for mixed states it is a challenge. The standard criteria that are generally applicable are based on the fidelity. Fidelity criteria set forth that any state $\rho$
 with GME dimension at most $r$ satisfies \cite{Malik2016, Cobucci2024}
 \begin{equation}\label{multifid}
	\bracket{\psi}{\rho}{\psi} \leq \max_{(P\lvert \bar{P})} \sum_{i=1}^{r} \lambda_i\left(\psi_P\right),
\end{equation}
where $\ket{\psi}_{1\ldots n}$ is some target state, $\psi_P=\tr_{\bar{P}}\left(\psi\right)$ is the reduced target state and $\{\lambda_i\}$ denote the eigenvalues ordered non-increasingly.

\subsection{BMM method for solving Problem 3}
We begin with outlining the main observation underpinning our construction of the SDP relaxation for the set of states with GME dimension at most $r$. To this end, let  $\rho$ be an $n$-partite and $d$-dimensional state with GME dimension $r$. Choose any bipartition $(P\lvert \bar{P})$. The state $\sigma_{(P\lvert \bar{P})}$ used in the decomposition \eqref{gmeeq} can in turn be decomposed into a mixture of pure states. Let $\ket*{\psi_{(P\lvert \bar{P})}}$ be any one of the pure states appearing in such a decomposition. Since the Schmidt number of $\sigma_{(P \lvert \bar{P})}$ is no larger than $r$, it follows that $\ket*{\psi_{(P\lvert \bar{P})}}$ can be supported on local Hilbert spaces of rank no larger than $r$. Hence, there exists projectors $\Pi_P$ and $\Pi_{\bar{P}}$, acting on the subsystems $P$ and $\bar{P}$ respectively, such that $\text{dim}(\Pi_P)=\text{dim}(\Pi_{\bar{P}})=r$ and
\begin{equation}\label{propgmedim}
	\Pi_P\otimes \Pi_{\bar{P}}\ket{\psi_{(P\lvert \bar{P})}}=\ket{\psi_{(P\lvert \bar{P})}}.
\end{equation}

To illustrate the main idea for the SDP relaxations, consider for simplicity the case when $\rho=\rho_{AB}$ is bipartite ($n=2$). Since there is only a single bipartition $P=\{A\}$ and $\bar{P}=\{B\}$.  Let $\rho=\sum_i p_i \psi_i$ be any pure-state decomposition of $\rho$ and let $\ket{\psi}_{AB}$ be any one of the pure states. The set of relevant operators in this scenario are $L=\{\Pi_A\otimes\mathds{1}, \psi\}$, where $\Pi_A$ is the projector of rank $r$ with the property \eqref{propgmedim}\footnote{We have ignored $\Pi_B$ since it is not needed for the SDP.}. Choosing the  relaxation level $S=L$, we obtain the BMM
\begin{equation}\label{GammaExampleGME}
	\Gamma=\left(\begin{array}{cc}
		\Pi_A\otimes\mathds{1} & \psi_{AB} \\
		\psi_{AB} & \psi_{AB}
	\end{array}\right) \ , 
\end{equation}
where we have used that  $\Pi_A^2=\Pi_A$, $\psi_{AB}^2=\psi_{AB}$ and $\Pi_A\otimes \mathds{1}\ket{\psi}=\ket\psi$. Furthermore, we  set $\tr\left(\Pi_A\right)=r$ as implied by the rank and projectivity of $\Pi_A$. Next, we can use the convexity of the program: if \eqref{GammaExampleGME} is SDP-feasible then any convex combination of it remains SDP-feasible. Our state of interest, namely $\rho_{AB}$, is such a convex combination. Therefore, we can without loss of generality replace $\psi_{AB}$ with  $\rho_{AB}$, whose  Schmidt number at most $r$. In \eqref{GammaExampleGME}, this leaves only the sub-block associated to $\Pi_A$ as a free variable. Using the Schur complement to simplify $\Gamma\succeq0$, we obtain the SDP relaxation 
\begin{equation}\label{eq:problem1_general}
	\begin{array}{ccc}
		\text{find}  & \Pi_A \\ [2ex]
		\text{s.t} & \Pi_A\otimes \mathds{1}\succeq  \rho_{AB}, \\[0.5ex]
		& \tr(\Pi_A)\leq r.
	\end{array}
\end{equation}

We  use the above as a springboard to the more pertinent multipartite regime ($n>2$).  Every bipartition $(P\lvert \bar{P})$ is considered separately and associated with a positive semidefinite matrix $\Gamma_{(P\lvert \bar{P})}$ of the type \eqref{GammaExampleGME}. By the decomposition \eqref{gmeeq} we consider the convex combination $\Gamma=\sum_{(P\lvert \bar{P})} q_{(P\lvert \bar{P})}\Gamma_{(P\lvert \bar{P})}$ of these matrices. It takes the form
\begin{align}\label{GammaMulti}
\Gamma&= \sum_{(P\lvert \bar{P})} q_{(P\lvert \bar{P})}\left(\begin{array}{cc}
\Pi_{P}\otimes\mathds{1}_{\bar{P}} & \sigma_{(P\lvert \bar{P})} \\
\sigma_{(P\lvert \bar{P})} & \sigma_{(P\lvert \bar{P})}
\end{array}\right).
\end{align}
Due to \eqref{gmeeq}, the three identical blocks can be fixed to the state $\rho_{1\ldots n}$. For the remaining  block, we define $\tilde{\Pi}_{P}=q_{(P|\bar{P})}\Pi_{P}$ and note that  $\tr\big(\tilde{\Pi}_P\big)=q_{(P|\bar{P})}r$ and  $0\preceq \tilde{\Pi}_P\preceq \frac{1}{r}\tr\left(\tilde{\Pi}_P\right)\mathds{1}$. We treat $\{\tilde{\Pi}_P\}$ as our variables, resulting in the SDP relaxation 
\begin{equation}\label{eq:SDP_GME_dim}
	\begin{aligned}
		\text{find} & & & \{\tilde{\Pi}_{P}\}_{(P\lvert \bar{P})}\\
		\text{s. t.} & & &  \sum_{(P\lvert \bar{P})}\tilde{\Pi}_P\otimes \mathds{1}_{\bar{P}} \succeq \rho_{1\ldots n}\\
		& & &   \sum_{(P|\bar{P})}\tr(\tilde{\Pi}_P)\leq r \\
		& & &  0\preceq  \tilde{\Pi}_P\preceq \frac{1}{r}\tr\left(\tilde{\Pi}_P\right)\mathds{1} \qquad \forall(P|\bar{P}) \ .
	\end{aligned}
\end{equation}
Note we can always select $P$ as the element in the bipartition such that $|P|\leq |\bar{P}|$. This makes the program more efficient but reduces the tightness of the resulting relaxation. If the program is infeasible, it implies that $\rho$ has a GME dimension larger than $r$.

\subsection{Results for Problem 3} 
We benchmark the SDP relaxation in three ways. First, we analytically derive its relation with fidelity criteria for GME-dimension. Second, we consider a well-known family of states as a case-study. Finally, we estimate the method's overall performance by applying it to randomly sampled states. 

The SDP relaxation for bounding GME-dimension is strictly stronger than  fidelity criteria, regardless of the choice of target state in Eq~\eqref{multifid}. Specifically, in Appendix~\ref{app:GME_ref1} we treat the bipartite case and prove that the SDP in Eq~\eqref{eq:problem1_general} is equivalent to the so-called singlet fraction, i.e.~fidelity estimation based on any state obtained by a local channel applied to maximally entangled state. In Appendix~\ref{app:GME_ref2}, we prove that such a target state leads to a strictly stronger fidelity criterion  than any other choice of target state, thereby implying that the bipartite version of the SDP relaxation can substitute for any possible fidelity criterion. More importantly, we extend the analysis to the multipartite regime in Appendix~\ref{app:GME_ref3}. There, we show that if the SDP relaxation in Eq~\eqref{eq:problem1_general} is feasible then the fidelity criterion \eqref{multifid} is satisfied for every choice of target state. This implies that the SDP is at least as powerful as fidelity criteria, even if we knew how to optimally choose the target state. As our case studies below will show, the advantage provided by the SDP is sizable. 

\begin{table}
	\begin{tabular}{|c|c|c|c|c|}
		\hline
		\diagbox{$d$}{$n$} & 3 & 4 & 5 & 6 \\ \hline
		2 & 0.5556 & 0.6127 & 0.5586 & 0.5676 \\ \hline
		3 & 0.6250 & 0.7391 & 0.7907 & 0.6633\\ \hline
		4 & $\ ***$ & 0.7460 & 0.7966 & 0.8326 \\ \hline
	\end{tabular}
	\caption{Upper bounds on the critical isotropic visibility for maximal GME dimension obtained for the $n$-partite and $d$-dimensional Dicke states. Bold numbers represent improvement over the fidelity criterion and other numbers represent equal results.  No Dicke state exists for $(d,n)=(4,3)$. }\label{Tab:dicke}
\end{table}

Consider a case study for a family of $n$-partite Dicke states. These are uniform superpositions of all bit-strings with Hamming weight $k$ \cite{Dicke}. We consider their generalisation to $d$-dimensional systems  \cite{Wei_2003, Hayashi2008}, defined as 
\begin{equation}
\ket{D^n_{\vec{k}}}=\frac{1}{\sqrt{C_{n,\vec{k}}}}\sum_{\text{perms}} \ket*{\underbrace{0\ldots 0}_{k_0} \hspace{0.5mm} \underbrace{1\ldots 1}_{k_1}\ldots\  \underbrace{d-1\ldots d-1}_{k_{d-1}} }\ , 
\end{equation}
where $C_{n,\vec{k}}=\frac{n!}{\Pi_{j=0}^{d-1} k_j}$ is the multinomial coefficient and the sum runs over all permutations of vectors in which the integer $j\in\{0,\ldots,d-1\}$ appears $k_j$ times. For an $n$-partite system, the vector $\vec{k}=(k_0,\ldots , k_{d-1})$ must satisfy $\sum_{i=0}^{d-1}k_i=n$.  We focus on the family corresponding to $k_j=\lceil \frac{n- j}{d}\rceil$. As a quantitative benchmark of performance we consider the mixture
\begin{equation}\label{rhov}
\rho_v=v\ketbra*{D^n_{\vec{k}}}{D^n_{\vec{k}}}+\frac{1-v}{d^n}\mathds{1},
\end{equation}
and use the SDP relaxation \eqref{eq:SDP_GME_dim} to place an upper bound on the critical value of $v\in[0,1]$ below which $\rho_v$  no longer has the maximal GME dimension ($r=d$). To strengthen our results, we use the formulation where $P$ is selected as the biggest element of the bipartition. To render the resulting problem computationally tractable, we further exploit the use of symmetries \cite{Rosset2019}. Specifically, we average each operator $\tilde{\Pi}_P$ over the group of subsystem permutations and local-level relabellings that leave the Dicke state invariant, thereby reducing the number of independent variables in the computation. The results are presented in \tabref{Tab:dicke} for up to six parties and up to four dimensions. All numbers represent an improvement over the fidelity criterion \eqref{multifid}.

\begin{table}[t]
	\centering
	\renewcommand{\arraystretch}{1.4}
	\begin{tabular}{|l||c|c|c|c|c|c|c|}
		\hline
		
		\multirow{2}{*}{$d$}
		& \multicolumn{3}{c|}{\textbf{a) \ n=3}}
		& 
		& \multicolumn{3}{c|}{\textbf{b) \ n=4}} \\
		\cline{2-4}\cline{5-8}
		& 	$\overline{v}^\text{fid}$ & $\overline{v}^\text{SDP}$ 
		& $\sigma(\delta)$ & &
		$\overline{v}^\text{fid}$ & $\overline{v}^\text{SDP}$ 
		& $\sigma(\delta)$ \\
		\hline
		4 & $0.9078$ & $0.8422$ & $0.0089$ && $0.8380$ & $0.7859$ & $0.0074$ \\
		\hline
		5 & $0.9202$ & $0.8727$ & $0.0061$ & & $0.8614$ & $0.8256$ & $0.0050$  \\
		\hline
		6 & $0.9280$ & $0.8915$ & $0.0052$ & & $0.8750$ & $0.8508$ & $0.0031$ \\
		\hline
		7 & $0.9351$ & $0.9058$ & $0.0034$ \\
		\cline{1-4}
		8 & $0.9397$ & $0.9158$ & $0.0033$ \\
		\cline{1-4}
		9 & $0.9447$ & $0.9244$ & $0.0026$ \\
		\cline{1-4}
	\end{tabular}
	\bigskip \\
	\centering
	\begin{tabular}{|c||c|c|c|c|}
		\hline
		\textbf{c) d = 3} &  n=3 & n=4 & n=5 & n=6 \\
		\hline
		$\overline{v}^\text{fid}$ & $0.8847$ & $0.8072$ & $0.7567$ & $0.7204$ \\
		\hline
		$\overline{v}^\text{SDP}$ & $0.7819$ & $0.7227$ & $0.6930$ & $0.6722$ \\
		\hline
		$\sigma(\delta)$ &  $0.0148$ & $0.0139$ & $0.0099$ & $0.0064$ \\
		\hline
	\end{tabular}
	\caption{Average critical isotropic visibility obtained employing the fidelity witness, $\overline{v}^\text{fid}$, and the SDP relaxation $\overline{v}^\text{SDP}$; $\sigma(\delta)$ defines the standard deviation of the gap $\delta=v^\text{fid}_i-v^\text{SDP}_i$. The results are presented for fixed number of parties, in Table $\textbf{a)}$ and $\textbf{b)}$ and fixed dimension, Table $\textbf{c)}$. All results are computed over $100$ randomly sampled pure states.}
	\label{tab:RES_gme}
\end{table}

The extent to which the SDP relaxation outperforms fidelity criteria depends on the specific state. To estimate this, as well as to show the computational efficiency of the method, we apply the cheapest formulation of \eqref{eq:SDP_GME_dim} to a number of randomly selected states. Specifically, for each choice of $(d,n)$, we sample 100 pure states, $\psi_i$, for $i=1,\ldots,100$, and compute a bound on the  critical isotropic visibility (we replace $\ket*{D^n_{\vec{k}}}$ with $\ket{\psi_i}$ in Eq~\eqref{rhov}) for detecting $d$-dimensional GME. The resulting bound obtained from the SDP relaxation is denoted $v_i^{\text{sdp}}$. We compare the results with those obtained from fidelity criterion \eqref{multifid} where we select $\psi_i$ as the target state and label the corresponding bound on the critical visiblity by $v^{\text{fid}}_i$. In parts a) and b) of Table~\ref{tab:RES_gme} we present the results for three- and four-partite cases for  $d=4,\ldots 9$.  The entries in the tables represent the average values, $\bar{v}^x=\frac{1}{100}\sum_i v_i^{x}$ for $x\in\{\text{sdp},\text{fid}\}$, and the standard deviation of the gap $\delta_i=v_i^{\text{sdp}}-v_i^{\text{fid}}$. We observe a significant advantage in all cases. The advantage decreases for larger $d$, but this is expected since the true value of the critical visiblity also is expected to increase monotonically toward unity with $d$. Notably, since the standard deviation also decreases with  $d$, the size of the advantage is more reliable in those cases for a randomly selected state. Similarly, in part c) of  Table~\ref{tab:RES_gme}, we present results for $n=3,4,5,6$ when $d=3$. Again, we see a sizable average gap in all cases, whose magnitude and standard deviation both decrease with $n$. Finally, we note that even larger systems are within reach of the SDP method, as indicated by the large number of repetitions we have been able to compute within reasonable time using a standard laptop for systems of size e.g.~six ququarts.

\section{Operational dimension of quantum preparation devices}\label{sec:abs_dim}

Determining the number of dimensions  required to describe a quantum device is a natural inquiry. It has  operational meaning in terms of the  necessary quantum control, the amount of  information it can transmit and what quantum information protocols it can support. For devices that generate a set of single-quantum systems, benchmarks of dimensionality have been studied intensively through the paradigm of device-independent quantum information \cite{Gallego2010, Ahrens2012, Hendrych2012}. This approach allows one to deduce the dimensionality without assuming any control of the measurement devices used extract the data on which the inference is based. However, the device-independent approach has a fundamental cost because many high-dimensional preparation devices are expected to be out of its reach. Also, it is many times more relevant to certify dimensionality in the standard paradigm of quantum information, namely when the measurement devices used in the process are assumed to be trusted. In such settings, one approach is to select a  basis with respect to which the number of coherent levels can be quantified \cite{Ringbauer2018}. A more fundamental approach is to adopt no priviledged basis and benchmark dimensionality in a basis-independent way \cite{AbsDimAlex}. However, no general method is available for the latter and results are limited to problems with a strong degree of symmetry. We show how to tailor BMM SDP relaxations to provide a general tool for certifying the  dimensionality needed to operate a quantum preparation device.

\subsection{Description of Problem 4}
Consider a quantum state preparation device that can on-demand emit any one of the states in the set $\mathcal{E}=\left\lbrace \rho_x \right\rbrace_{x=1}^m$, where each state lives in a $d$-dimensional Hilbert space. Although the physical dimension is $d$, one can ask whether the states $\mathcal{E}$ can be simulated by using classical pre- and post-processing of other hypothetical quantum  devices which are limited to operating Hilbert spaces of dimension $r<d$. If affirmative, quantum control of $r$-dimensional systems is sufficient to generate $\mathcal{E}$. Therefore, the smallest value of $r$ can be viewed as the operational dimension of the quantum preparation device.  Formally, let $\{\Pi_{\lambda}\}$ denote any set of projectors onto $r$-dimensional subspaces and let $\{q(\lambda)\}$ be a probability density. The operational dimensionality of $\mathcal{E}$ is defined as  \cite{AbsDimAlex}
\begin{align}\label{eq:abs_dim}
	r_Q(\mathcal{E})\equiv \min_{\{q\}, \{\sigma\}}\lbrace &r : \rho_x=\int d\lambda\ q(\lambda)\sigma_{x,\lambda},\\
		&\text{where} \quad \exists\Pi_\lambda \quad \text{s.t.} \quad \Pi_\lambda^2=\Pi_\lambda, \nonumber \\
		&\tr(\Pi_\lambda)=r, \quad \sigma_{x,\lambda}=\Pi_\lambda\sigma_{x,\lambda}\Pi_\lambda \ \ \forall \ \ x \rbrace \ . \nonumber
\end{align}
Our task is to find a general and efficient way to bound the set of $\mathcal{E}$ for any given value of the operational dimensionality $r_Q$.

\subsection{BMM method for solving Problem 4}
Since the set $\mathcal{E}$ is defined for a given Hilbert space of dimension $d$, we design our BMM such that its blocks have size $d\times d$. To this end, we choose the map  $\Theta=\text{id}$ in Eq~\eqref{eq:block_mat}. To motivate the construction of the SDP relaxation,  let us momentarily consider that we fix the value of the classical random variable $\lambda$ appearing in \eqref{eq:abs_dim}. The set of operators appearing in the problem becomes $L=\{\Pi_\lambda,\sigma_{x,\lambda}\}$. Once we select the monomial list $S$ we obtain a BMM which we label $\Gamma_\lambda$. We can use the same argument as discussed for Problem 3 to account for the convexification present  in \eqref{eq:abs_dim}. Specifically, once our SDP-constraints on $\Gamma_\lambda$ are identified, we can define a new BMM as $\Gamma=\int d\lambda q(\lambda)\Gamma_\lambda$  then work directly with $\Gamma$ in our SDP relaxation. As an easy illustration, consider the example of  a set with $m=2$ preparations, $\{\rho_1,\rho_2\}$, and a BMM corresponding to $S=\{L\}\cup \{\mathds{1}\}$. The BMM becomes
\begin{equation}
	\Gamma=\!\!\int\! d\lambda q(\lambda)
\begin{pmatrix}
\mathds{1} & \Pi_\lambda & \sigma_{1,\lambda} & \sigma_{2,\lambda} \\
& \Pi_\lambda & \sigma_{1,\lambda} & \sigma_{2,\lambda} \\
& & \sigma_{1,\lambda} & Q_\lambda \\
& & & \sigma_{2,\lambda}
\end{pmatrix} =	
	\begin{pmatrix}
		\mathds{1} & \Pi & \rho_1 & \rho_2 \\
		& \Pi & \rho_1 & \rho_2 \\
		& & \rho_1 & Q \\
		& & & \rho_2 
	\end{pmatrix} \ ,
\end{equation}
where in the first step we have used that $\Pi_\lambda^2=\Pi_\lambda$, $\Pi_\lambda\sigma_{x,\lambda}=\sigma_{x,\lambda}$ and that $\sigma_{x,\lambda}$ can be assumed pure without loss of generality. Due to the rank of $\Pi_\lambda$, we  impose  $\tr\left(\Pi_\lambda\right)=r$. In the above, we have defined  $Q_\lambda=\sigma_{1,\lambda}\sigma_{2,\lambda}$ as an unconstrained variable. Both the positivity of the BMM and the constraints on each block are preserved over the convexification: the variable $\Pi=\int d\lambda q(\lambda) \Pi_\lambda$ is Hermitian and satisfies $\tr(\Pi)=r$, while the variable $Q=\int d\lambda q(\lambda) Q_\lambda$ remains unconstrained. Moreover, the convexification yields the elements of $\mathcal{E}$ because $\rho_x=\int d\lambda q(\lambda)\sigma_{x,\lambda}$. Thus, these constraints together with $\Gamma\succeq 0$ define an SDP relaxation of those $\mathcal{E}$ that can be simulated using quantum preparation devices of dimension no more than $r$.  By extending this to larger $\mathcal{E}$ and sometimes also to higher relaxation levels, we obtain a general SDP relaxation method where the states to be simulated, $\{\rho_x\}_x$, appear as specific blocks and the trace of $\Pi$ determines the choice of $r$.

\subsection{Results for Problem 4}
We showcase the performance of the method in two ways. Firstly, we consider a family of states studied in \cite{AbsDimAlex} and show that the SDP method provides rigorous bounds where previously only numerical estimates were known. Secondly, we consider the action of a phase-damping channel applied to a device that generates eigenstates of the Fourier basis and we determine bounds on the critical amount of damping needed to reduce the operational dimensionality. In both cases, we can apply the method to systems whose dimension goes well above the few simplest cases while obtaining useful  results.

For the former, consider the first $d-1$ states of the computational basis, namely $\{\rho_x=\ketbra{x}\}_{x=0}^{d-2}$, and the uniform superposition state $\rho_{d-1}=\ketbra{\phi}$ where $\ket{\phi}=(\ket{0}+\ldots+\ket{d-1})/\sqrt{d}$. To each of them we apply isotropic noise with visibility $v\in[0,1]$, leading to the mixed state 
\begin{equation}
\rho_x(v)=v\rho_x+(1-v)\frac{\mathds{1}}{d}.
\end{equation}
The set of states under consideration is therefore  $\mathcal{E}=\{\rho_x(v)\}_{x=0}^{d-1}$. Our goal is to determine upper bounds on the critical value of $v$ for a simulation using $r$-dimensional quantum devices.  For this we use the SDP method at the first-level of the relaxation hierarchy ($K=1$) and use  $v$ as the objective function. Note that the computational cost is constant for the different choices of $r$ since it enters only by varying a linear constraint in the SDP. The upper bounds on the critical visibility for $d=8$ are presented in \tabref{tab:merged_abs_dim}~\textbf{a)} together with the numerical lower-bounds reported in \cite{AbsDimAlex} by a brute-force search method. We see that the gap between the bounds is small, especially for small $r$. This attests to the accuracy of our method, especially in view of the remark in \cite{AbsDimAlex} that the lower bounds are suboptimal. In addition, while this example focuses on $d=8$, we have been able to compute bounds for this problem up $d=15$ in reasonable time using a standard laptop.

Let us now consider a different case study. Consider the $d$-dimensional Fourier basis  $\ket{f_x}=\frac{1}{\sqrt{d}}\sum_{j=0}^{d-1}\omega^{xj}\ket{j}$ with $\omega=e^{\frac{2\pi i}{d}}$. All states in this basis have uniform population and differ only by a sequence of phases. We send each state through a phase-damping channel. The channel's Kraus operators are given by $K_0=\ket{0}\bra{0}+\sqrt{1-\gamma}\sum_{l=1}^{d-1}\ketbra{l}$ and $K_j=\sqrt{\gamma}\,\ket{j}\bra{j}$ for $j=1,\dots,d-1$, where $\gamma\in[0,1]$ is the damping parameter. Hence, the mixed output states become  $\rho_x(\gamma)=\sum_{i=0}^{d-1} K_i\ketbra{f_x}K_i^\dagger$. We will bound from below the critical value of $\gamma$ for set $\mathcal{E}=\{\rho_x(\gamma)\}_x$ admitting  a simulation with $r$-dimensional quantum devices.  However, in the SDP we cannot treat $\gamma$ as a variable due to its nonlinearity in $\Gamma$. Therefore, for a given $r$, we fix $\gamma$ and gradually increase it until the SDP (at first level) becomes feasible. Our bound then corresponds to the largest $\gamma$ in the sequence for which the program is infeasible. The results for $d=8$ are displayed in \tabref{tab:merged_abs_dim}~\textbf{b)}. As expected, imposing more structure on the noise model, improves the robustness of the dimensionality certificate. For instance, if we consider isotropic noise for this example, the maximum dimension can be certified only up to a noise fraction $(1-v)\leq0.1429$, whereas under phase damping the corresponding threshold is $\gamma\leq0.1623$.

\begin{table}[t]
\renewcommand{\arraystretch}{1.1}
\centering
\begin{tabular}{|l||c|c|c|c|}
\hline
\multirow{3}{*}{$r$}
  & \multicolumn{2}{c|}{\textbf{a) \ Comp.~basis + mixed}}
  & 
  & \textbf{b) \ Fourier basis} \\
\cline{2-3}\cline{5-5}  
& \multicolumn{2}{c|}{Isotropic visibility $v$}
& 
& Phase damping $\gamma$ \\
\cline{2-3}\cline{5-5}
 & Lower-bound & Upper-bound 
 & 
 & Lower-bound  \\
\hline
2 & 0.1537 & 0.1579 &  & 0.9097  \\
\hline
3 & 0.3099 & 0.3158 &  & 0.7759  \\
\hline
4 & 0.4647 & 0.4737 &  & 0.6311  \\
\hline
5 & 0.6133 & 0.6316 &  & 0.4788 \\
\hline
6 & 0.7525 & 0.7894 &  & 0.3221 \\
\hline
7 & 0.8800 & 0.9369 &  & 0.1623 \\
\hline
\end{tabular}
\caption{Bounds for certifying operational dimensionality larger than $r$ for \textbf{a)} an ensemble of seven computational basis elements and the $8$-dimensional uniform superposition state under isotropic noise, and \textbf{b)} the $8$-dimensional Fourier basis when passed through a phase-damping channel elements. In case \textbf{a)} we compare our upper-bounds with the lower-bounds reported in \cite{AbsDimAlex}.}
\label{tab:merged_abs_dim}
\end{table}

\section{Uncertainty relations}\label{sec:uncertainty}

Uncertainty relations impose limitations on how precisely one can predict the outcomes of  non-commuting observables: better knowledge of one limits the knowledge of another. Such relations provide a powerful tool for example in quantum metrology and in quantum cryptography \cite{Coles2017}. A simple example for qubits is the uncertainty relation $\expect{\sigma_X}^2+\expect{\sigma_Y}^2+\expect{\sigma_Z}^2\leq 1$, where the observables anticommute. Such observables are broadly relevant, for instance in quantum error correction where anticommutation with stabilizer elements  is used to detect errors. Uncertainty relations based on sums-of-squares for anticommuting observable have for instance been used as  key ingredients in criteria for entanglement \cite{Toth2005, Hansenne2022}, entropic uncertainty relations \cite{Wehner2008, Niekamp2012} and Bell monogamy relations \cite{Kurzy2011}. More generally, when only some observables  anticommute, dedicated SDP methods have been developed for bounding the sum-of-squares  \cite{Gois2023, Moran2024}.  However, as pointed out in \cite{Gois2023}, it is relevant from a practical aspect to consider observables that only ''almost'' anticommute, i.e.~to make the uncertainty relations robust to imprecisely calibrated devices. We show that the BMM SDP relaxations can account for constraints associated with  such partially relaxed anticommutation relations and that it can provide tight bounds for a sizable number of observables.

\subsection{Description of Problem 5}
Ref~\cite{Gois2023} introduced a general form for uncertainty relations  based on sums-of-squares of expectation values for observables with anticommutation properties. Specifically, consider any set of observables $\{O_i\}_{i=1}^n$ with eigenvalues $\pm 1$ and the quantity
\begin{equation}\label{eq:beta_uncertainty}
	\beta=\sup_\rho\sum_{i=1}^n \expect{O_i}_\rho^2.
\end{equation}
The anticommutation relations are described by a graph: each observable is a vertex and two vertices are connected by an edge if  the observables anticommute. For any anticommutation graph, an SDP-computable upper bound on $\beta$ was given in \cite{Gois2023}, which was later improved by a hierarchy of SDPs in \cite{Moran2024}. 

However, these SDPs do not apply when we are interested in making the uncertainty relation robust to calibration errors. Ref~\cite{Gois2023} considered a model where if the vertices $i$ and $j$ are connected ($i\sim j$), the operator norm of the anticommutator  satisfies  
\begin{equation}\label{anticomrel}
	-\eta_{ij}\mathds{1}\preceq \{O_i,O_j\} \preceq \eta_{ij}\mathds{1},
\end{equation} 
for some small parameters $\eta_{ij}\geq 0$ representing the calibration error. This complicates the computation  in \eqref{eq:beta_uncertainty} because the supremum now includes also all observables compatible with the condition \eqref{anticomrel} for every pair of connected vertices in the graph. An analytical bound is available if one additionally assumes that the graph structure is preserved. However, this departs from the limited control motivation underlying this framework. The task is therefore to systematically determine accurate bounds on $\beta\left(\{\eta_{ij}\}\right)$ without imposing additional constraints.

\subsection{BMM method for solving Problem 5} 
To construct an appropriate BMM on which the SDP relaxations are based, we must first choose the operator list $L$. In the four previous problems we have chosen $L$ based on the operators actually appearing in the problem, but this is not viable here because it does not allow us to access in an SDP-compatible way the quadratic terms $\expect{O_i}^2$ appearing in our objective \eqref{eq:beta_uncertainty}. Instead, we choose to append to the list with a so-called scalar extension  (see e.g.~\cite{Pozas2019}). Specifically, we select  $L=\{\rho, O_i,\tilde{O}_i\}$, where we have defined $\tilde{O}_i=\expect{O_i}O_i$. The reason for including also $\tilde{O}_i$ is that it will let us express the squared expectation values in \eqref{eq:beta_uncertainty} as linear functions over blocks in the BMM. Furthermore, since $O_i$ are $d$-dimensional observables, we build the BMM \eqref{eq:block_mat} with the identity map $\Theta=\text{id}$.  To illustrate it, consider the first relaxation level ($K=1$), corresponding to the monomial list $S=\{L\}\cup \{\mathds{1}\}$. The BMM becomes
\begin{equation}
	\Gamma=\begin{pmatrix}
		\mathds{1} & \rho &\{O_i\} & \{\expect{O_i}O_i\} \\[0.5ex]
		&  \rho & \{\rho O_i\}  & \{\expect{O_i}\rho O_i \}  \\[0.5ex]
		& & \{O_jO_i\} & \{\expect{O_i}O_jO_i\}  \\[0.5ex]
		& & & \{\expect{O_i}\expect{O_j}O_jO_i\} 
	\end{pmatrix},
\end{equation}
where we have without loss of generality set $\rho^2= \rho$.  The terms relevant for computing \eqref{eq:beta_uncertainty} are $\expect{O_i}^2$ and these are obtained from tracing the blocks $\Gamma_{\rho,\tilde{O}_i}=\expect{O_i}\rho O_i$. Also, a number of constraints  between the blocks  is obtained from the property $O_i^2=\mathds{1}$. Moreover, the relaxed anticommutation relations \eqref{anticomrel} can be imposed as operator inequalities over the relevant blocks in the array $\{O_jO_i\}$. Furthermore, notice  that many other types of observable constraints can be incorporated into the method: we can for instance impose any semidefinite conditions on the  observable products $O_jO_i$, or on the individual observables $O_i$. Finally, we point out that also constraints such as $\tr(\Gamma_{\rho,O_i})^2=\tr(\Gamma_{\rho,\tilde{O}_i})$ are relevant, but we are unable to impose them since the nonlinearity is incompatible with SDP. Putting all this together gives the SDP relaxation 
\begin{align}\nonumber\label{BMMprob5}
	\max_\Gamma & \quad \sum_{i=1}^n \tr(\Gamma_{\rho, \tilde{O}_i})  \\\nonumber
	\text{s.t.} & \quad -\eta_{ij}\mathds{1}\preceq \Gamma_{O_i,O_j}+\Gamma_{O_j,O_i}\preceq \eta_{ij}\mathds{1} \qquad \forall  i\sim j  \\\nonumber
	& \quad \Gamma_{O_i,O_i}=\mathds{1},\\\nonumber
	&  \quad \Gamma_{O_i,\tilde{O}_i}=\tr\big(\Gamma_{\rho,O_i}\big)\mathds{1},\\\nonumber
	& \quad \Gamma_{\tilde{O}_i,\tilde{O}_i}=\tr\big(\Gamma_{\rho,\tilde{O}_i}\big)\mathds{1},  \\\nonumber
	& \quad \tr\left(\Gamma_{\mathds{1},\rho}\right)=1,\\
	& \quad \Gamma\succeq 0 \ . 
\end{align}
When evaluating this program, we can restrict ourselves to real-valued BMMs, because if a complex-valued  $\Gamma$ is feasible then  $(\Gamma+\Gamma^\dagger)/2$ is also feasible and it leaves the objective unchanged. The above SDP represents only the first relaxation level, but we will find that with only a small extension of the monomial list the method can provide optimal bounds.

\subsection{Results for Problem 5}
To showcase the performance of the SDP relaxations, we  focus on anticommutation graphs in the shape of a ring, i.e.~the graph is an $n$-cycle where observable $O_{i}$ is connected to the observables $O_{i-1}$ and $O_{i+1}$ (where $O_{n+1}=O_1$ and $O_{0}=O_n$).  For instance, for $n=3,4,5$ the graph is a triangle, square and a pentagon respectively. We will consider it for $n=3,4,\ldots,21$ traceless qubit observables. Hence,  the condition \eqref{anticomrel} is satisfied for every observable when considering its two  neighbours. While we can address arbitrary parameter sets $\{\eta_{ij}\}$ with unchanged computational cost, we choose for simplicity of presentation that $\eta_{ij}=\eta$.

Our computations are based on an intermediate relaxation level: we extend the first level by adding all monomials of the form $\tilde{O}_{i+1}O_{i+1}O_i$.   For every choice of $n$, we find that the resulting upper bound on $\beta_n(\eta)$ scales linearly with the parameter $\eta$, 
\begin{equation}\label{eq:linear_beta}
\beta_n(\eta) = \beta_n(0)+\alpha_n \eta,
\end{equation}
where $\beta_n(0)$ is the bound under exact anticommutation. When $n$ is even, one has $\beta_n(0)=n/2$, which is saturated e.g.~by assigning $O_i=\sigma_X$ for odd $i$ and $O_i=\sigma_Z$ for even $i$ and choosing $\rho=\frac{1}{2}\left(\mathds{1}+\frac{1}{\sqrt{2}}\left(\sigma_X+\sigma_Z\right)\right)$. For even $n$, we find from the SDPs that the correction term is given by $\alpha_n=n/4$. The numbers are less elegant when $n$ is odd; we provide them case-to-case in \tabref{Tab:uncertainty}. For every $n$ considered, we have been able to match the upper bounds up to solver precision with lower bounds obtained by brute force optimisation.

\begin{table}[t]
	\centering
	\renewcommand{\arraystretch}{1.4}
	
	\begin{minipage}[t]{0.15\textwidth}
		\centering
		\begin{tabular}{|l|l|l|}
			\hline
			$n$  & $\beta_n(0)$    & $\alpha_n$                  \\ \hline
			$3$  & $1$          & $1$                         \\ \hline
			$5$  & $\sqrt{5}$   & $\frac{5-\sqrt{5}}{2}$      \\ \hline
			$7$  & $3.318$      & $1.841$                     \\ \hline
			$9$  & $4.360$      & $2.320$                     \\ \hline
			$11$ & $5.38630$    & $2.807$                     \\ \hline
		\end{tabular}
	\end{minipage}%
	\begin{minipage}[t]{0.15\textwidth}
		\centering
		\begin{tabular}{|l|l|l|}
			\hline
			$n$  & $\beta_n(0)$    & $\alpha_n$                  \\ \hline
			$13$ & $6.404$      & $3.298$                     \\ \hline
			$15$ & $7.417$      & $3.791$                     \\ \hline
			$17$ & $8.427$      & $4.286$                     \\ \hline
			$19$ & $9.435$      & $4.783$                     \\ \hline
			$21$ & $10.441$     & $5.279$                     \\ \hline
		\end{tabular}
	\end{minipage}
	
	\caption{Tight bounds for $\beta_n(\eta)$ (see Eq~\eqref{eq:linear_beta}) for almost anticommuting qubit observables corresponding to an $n$-cycle graph.}
	\label{Tab:uncertainty}
\end{table}

Finally, we point out that the obtained uncertainty relations  directly imply entanglement witnesses for anticommuting measurements that are  robust to calibration errors. To see this connection, one needs only to consider an entanglement witness of the form $\sum_{i=1} \expect{A_i\otimes B_i}$, where $\{A_i\}_i$ and $\{B_i\}$ could each have their own anticommutation graph and their own parameters $\{\eta_{ij}\}$ and $\{\eta'_{ij}\}$ respectively. Its maximum over separable states is achieved with a product state, for which the Cauchy-Schwarz inequality implies
\begin{equation}
	\sum_{i=1} \expect{A_i\otimes B_i}\leq \sqrt{\beta_A\left(\{\eta_{ij}\}\right) \beta_B\left(\{\eta'_{ij}\}\right)}.
\end{equation}
A violation implies that the state is entangled.

\section{Discussion}\label{sec:conclusions}
We have introduced a methodology for computing outer approximations to quantum correlations through hierarchies of semidefinite programs based on suitably tailored block-matrices. It is designed to be useful for treating problems featuring various algebraic, semidefinite and linear constraints over fully or partially characterised Hilbert spaces. By adapting it to five separate quantum information problems, each with their own physical restrictions, we have showcased a variety of different constraints  that can be addressed by the methodology. Examples include optimisation in a given Hilbert space over separable states, fidelity-constrained measurements, bounded operator norms and various dimension constraints, as well as optimisation in uncharacterised Hilbert spaces over states subject only to subspace constraints.  Many other types of constraints can be similarly  integrated.

The results obtained for the five problems represent different types of progress. For Problems 1, 2, 4 and 5 the SDP relaxations designed here are to our knowledge the first general method to address the respective problem. In the former case, it let us to go beyond the limitations of available analytical results and treat more experimentally relevant situations. In the next two cases, we obtained (at worst) nearly optimal results from computations that are easily performed  beyond the few lowest Hilbert space dimensions using standard laptops. In the final case, we obtained optimal results easily scalable to over 20 observables.  For Problem 3, we proved that the SDP relaxations are strictly stronger than the most commonly used analytical criteria  and we showed that the advantages are both significant in size and easy to compute for systems that are too large for  established  compter-based methods.

In general, the block structure is associated with an increase in the size of the moment matrix and therefore also with the computational cost of evaluating the SDPs. For a fixed monomial list of size $|S|$, replacing scalar entries with $d\times d$ blocks makes the matrix dimension grow as $d|S|$. Our applications exemplify two extremes of the trade-off between $d$ and $|S|$. Problem 3 has a minimal monomial list ($|S|=2$) and we can consequently analyse very large blocks (e.g.~size 4096). In contrast, Problem 5 has a moderate block dimension and we can consequently analyse a large monomial list  (e.g.~21 observables, computed in just $140$ seconds). Problems 1, 2 and 4 represent intermediate cases. For these, we are able to consider monomial lists sufficiently large to yield useful results while simultaneously having block dimensions in the double-digit size. For instance, for Problem 4 we almost double the treatable dimension compared to the numerical method of Ref~\cite{AbsDimAlex} and for Problem 3 we can handle significantly larger entangled systems than the numerical method of Ref~\cite{GMECobucci}. This attests to the practicality of the methodology and that is further strengthened  by the fact that all computations in this work have been performed using only a standard desktop computer.

\subsection{Relation to previous literature}

Our focus in this article has been the advances made possible by the proposed SDP relaxation methodology. Nevertheless, it remains relevant to discuss how some purpose-specific SDP relaxation hierarchies reported in the earlier literature fit into the  framework discussed here. Below, we will discuss three different examples of how such results emerge as special cases of our discussion.

Firstly, the NPA-type moment matrix that is standard in for example quantum nonlocality \cite{Navascues2007, Navascues2008} can be obtained as a limiting case. It  corresponds to the extreme case in which blocks comprising the block moment matrix are reduced to scalars. Specifically, we need only to choose in  Eq~\eqref{eq:block_mat} the map $\Theta$ as an inner product map for an unknown pure state $\psi$, namely $\Theta(X)=\tr\left(X\psi\right)$. 

Secondly, a well-studied problem in the literature is that of bounding quantum correlations in Bell scenarios or prepare-and-measure scenarios when states and measurements are restricted to a known Hilbert space. Some well-performing SDP methods methods are limited to heuristic sampling techniques \cite{NavascuesVertesi2015, Navascues2015b, Rosset2019}. Other SDP methods that are provably converging do not perform well at practically computable levels \cite{Navascues2014b, Jee2021}. Also, some methods that perform reasonably well and are free from sampling heuristics do not converge \cite{jef_almost_qudit}. Meanwhile, addressing this problem appears particularly natural with our approach, since the dimension can be imposed through the block-size of the SDP matrix. As we show in Appendix~\ref{app:dimension_hierarchy} the most immediate adaption of the methodology to address the specific problem of dimension-constrained quantum correlations ultimately reduces to the SDP relaxation hierarchy put forward in \cite{jef_almost_qudit}.

Thirdly, another type of quantum correlations is known as quantum steering. It is a formalisation of the Einstein-Podolsky-Rosen paradox and reveals whether a local variable can model the update made to a system by remotely measuring another system with which it is entangled \cite{Wiseman2007}. While this problem can be solved exactly as a single SDP \cite{Cavalcanti2017}, the same is no longer possible when the bar is raised to detecting the so-called genuine dimensionality of steering \cite{Designolle2021}. Dedicated SDP methods have been developed for this latter problem: one approach is less useful for practical computer resources but it is known to converge to the exact solution \cite{Gois2023}, while another approach is more practically handy but its convergence is unknown \cite{SDP_Steering}. In Appendix~\ref{app:steering} we show that the latter approach can be viewed as yet another problem-specific application of the overarching methodology of this article. 

We remark that other SDP relaxations relying on a block matrix structure appear in the literature for specialised settings. In the context of steering, a technique to quantify entanglement negativity compatible with an observed assemblage have been introduced in \cite{Pusey2013}. In a different direction, a convergent hierarchy for extended non-local  and monogamy-of-entanglement games was proposed in \cite{johnston2016extended}.

\subsection{Outlook}
The above discussion also underlines that the convergence properties of the SDP relaxation methodology cannot be addressed in general, but should instead be considered on a more problem-specific basis. Notably, it cannot only be addressed on the level of the map $\Theta$, because many different quantum correlation problems may be associated to the same $\Theta$ but their convergence may be decided by the problem-specific constraints and how the specific SDP relaxation is designed. Understanding the convergence properties in detail is an avenue for future work. 

We believe the proposed methodology for tailoring SDP relaxations to quantum correlation problems has applications well-beyond the five different problems discussed explicitly in this article. For example, the method may be potentially useful for benchmarking protocols in various forms of quantum key distribution and quantum random number generation, where it may be able to systematically take relevant physical imperfections into account (see e.g.~\cite{Woodhead2013, Margarida2020}).  Another potential direction is the falsification of alternative physical models for quantum devices, for example projective models for measurements \cite{Oszmaniec2017} or superposition-free models for quantum states  \cite{Cobucci2026}. More generally, these methods may find applications outside of physical science, in problems concerning non-commutative polynomial optimisation.

\begin{acknowledgments}
We thank Alexander Bernal and Miguel Navascu\'es for feedback. This work is supported by the Wenner-Gren Foundations, by the Knut and Alice Wallenberg Foundation through the Wallenberg Center for Quantum Technology (WACQT), the Swedish Research Council under Contract No.~2023-03498 and  the Swedish Foundation for Strategic Research.
\end{acknowledgments}

\section*{Code availability}
All codes used to generate the results in this study are available at \cite{githubRep}. The repository include the Julia library developed for implementing the BMM SDP relaxation and the implementation of the five problems presented above.

\onecolumngrid
\appendix

\section{SDP criterion vs fidelity criteria for multipartite high-dimensional genuine multipartite entanglement}\label{app:GME_ref}

\subsection{Bipartite systems: SDP criterion is equivalent to singlet fraction criterion}\label{app:GME_ref1}
On the one hand, consider our SDP \eqref{eq:SDP_GME_dim} from the main text. When considering only bipartite states ($n=2$), it simplifies to
\begin{equation}\label{eq:SDP_SN}
\begin{aligned}
\text{find} & & & \Pi \\
\text{s. t.} & & &  \Pi\otimes \mathds{1} \succeq \rho_{AB}\\
& & &   \tr(\Pi)\leq r\quad.
\end{aligned}
\end{equation}
Note that $\Pi\succeq 0$ is implied and that we can without loss of generality restrict to $\Pi \preceq \mathds{1}$.

On the other hand, consider the singlet fraction, $\text{SF}$, which is defined as
\begin{equation}
\text{SF}\left(\rho_{AB}\right) = \max_{\Lambda} \bracket{\phi^+}{(\Lambda\otimes\mathds{1})[\rho_{AB}]}{\phi^+},
\end{equation}
where $\ket{\phi^+}=\frac{1}{\sqrt{d}}\sum_{i=0}^{d-1}\ket{ii}$ is the maximally entangled state and the maximisation is over any completely positive trace-preservning map $\Lambda$. Since this can equivalently be written in terms of the dual map $\Lambda^\dagger$ as 
\begin{equation}
\text{SF}\left(\rho_{AB}\right) = \max_{\Lambda} \tr\left(\left(\Lambda^\dagger\otimes\mathds{1}\right)[\phi^+]\rho_{AB}\right),
\end{equation}
it can be viewed as a fidelity criterion for any state reachable from the maximally entangled state by a unital channel. Hence, for Schmidt number at most $r$, we have  
\begin{equation}\label{SFfid}
\text{SF}\left(\rho_{AB}\right)\leq \frac{r}{d}.
\end{equation}
We now show that this inequality for the singlet fraction is equivalent to the SDP criterion in \eqref{eq:SDP_SN}. 

For the proof, we  use Choi representation to write the singlet fraction as the SDP
\begin{equation}
\begin{aligned}
\text{max} & & & \tr\left(\rho_{AB} \eta_{AB}\right) \\
\text{s. t.} & & &  \tr_B\left(\eta_{AB}\right)=\frac{\mathds{1}}{d}\\
& & & \eta_{AB}\succeq 0 \qquad  \ .
\end{aligned}
\end{equation}
The dual program reads
\begin{equation}
\begin{aligned}
\text{min} & & & \frac{1}{d}\tr\left(Z\right) \\
\text{s. t.} & & &  Z\otimes\mathds{1}-\rho_{AB} \succeq 0 \quad.
\end{aligned}
\end{equation}
Due to strong duality of SDP, both the primal and the dual are equal to the singlet fraction. Imposing the condition \eqref{SFfid} means $\frac{1}{d}\tr\left(Z\right) \leq \frac{r}{d}$, namely $\tr\left(Z\right)\leq r$. Hence, we have recovered the SDP condition \eqref{eq:SDP_SN}. Thus, these criteria are equivalent.

\subsection{Bipartite systems: fidelity with maximally entangled states is sufficient}\label{app:GME_ref2}
One may consider that for some states $\rho_{AB}$ it is advantageous for fidelity-based Schmidt number witnessing to use a target state that is not maximally entangled. We prove that this is not the case. In other words, we show that the optimal target state is always maximally entangled (up to a local unital channel). Taken together with the result of the previous section, it follows the the BMM SDP relaxation for Schmidt number detection is equivalent to the fidelity-based criteria under arbitrary target states.

To prove the result, consider a general target state $\ket{\psi}$. The fidelity criterion becomes
\begin{equation}\label{fidcond}
\bracket{\psi}{\rho_{AB}}{\psi} \leq \sum_{i=1}^r \lambda_i\left(\psi_A\right),
\end{equation}
where the eigenvalues $\lambda_i$ are ordered decreasingly. We will now prove that if the SDP \eqref{eq:SDP_SN} is feasible, then  \eqref{fidcond} is satisfied. 

For this purpose, assume that \eqref{eq:SDP_SN} is feasible. Since $\Pi\otimes\mathds{1}-\rho_{AB}\succeq 0$ it holds in particular that
\begin{equation}\label{stepeq}
0\leq \bracket{\psi}{\Pi\otimes\mathds{1}-\rho_{AB}}{\psi} = \tr\left(\Pi \psi_A\right) -\bracket{\psi}{\rho_{AB}}{\psi}.
\end{equation}
Write $\Pi$ and $\psi_A$ is their spectral decomposition as
\begin{align}
& \Pi=\sum_{i=1}^d \mu_i \ketbra{\mu_i}{\mu_i}, \qquad \text{and} \qquad  \psi_A=\sum_{i=1}^d  \lambda_i \ketbra{\lambda_i}{\lambda_i}  
\end{align}
where we have ordered both $\mu_i$ and  $\lambda_i$ decreasingly. From \eqref{eq:SDP_SN} we have that 
\begin{equation}\label{condsmu}
0\leq \mu_i \leq 1 \qquad \text{and} \qquad  \sum_{i=1}^d \mu_i \leq r.
\end{equation}
This leads to 
\begin{equation}
\tr\left(\Pi \psi_A\right) = \sum_{i,j=1}^d \lambda_i \mu_j \left|\braket{\lambda_i}{\mu_j}\right|^2 = \sum_{i,j=1}^d \lambda_i \mu_j T_{ij}
\end{equation}
where we have defined the doubly stochastic matrix $T_{ij}=\left|\braket{\lambda_i}{\mu_j}\right|^2$. Since the eigenvalues are ordered decreasingly and since $T$ is doubly stochastic, the right-hand-side is maximised by choosing $T_{ij}=\delta_{ij}$. This gives 
\begin{equation}
\tr\left(\Pi \psi_A\right) \leq \sum_{i=1}^{d} \lambda_i \mu_i \leq \sum_{i=1}^{r} \lambda_i,
\end{equation}
where in the second step we used the conditions \eqref{condsmu} to optimally choose $\mu_1=\mu_2=\ldots=\mu_r=1$ and the remaining $\mu_i=0$.  Inserting this into \eqref{stepeq} we obtain
\begin{equation}
0\leq \sum_{i=1}^{r} \lambda_i -\bracket{\psi}{\rho}{\psi},
\end{equation}
which is the fidelity criterion.

\subsection{Multipartite systems: SDP criterion is stronger than any fidelity criterion}\label{app:GME_ref3}
On the one hand, consider again the SDP from the main text for $n$-partite systems of local dimension $d$, 
\begin{equation}\label{eq:SDP_GME_dim2}
	\begin{aligned}
		\text{find} & & & \{\tilde{\Pi}_{P}\}_{(P\lvert \bar{P})}\\
		\text{s. t.} & & &  \sum_{(P\lvert \bar{P})}\tilde{\Pi}_P\otimes \mathds{1}_{\bar{P}} \succeq \rho_{1\ldots n}\\
		& & &   \sum_{(P|\bar{P})}\tr(\tilde{\Pi}_P)\leq r \\
		& & &  0\preceq  \tilde{\Pi}_P\preceq \frac{1}{r}\tr\left(\tilde{\Pi}_P\right)\mathds{1} \qquad \forall(P|\bar{P}) \ .
	\end{aligned}
\end{equation}
On the other hand, consider the general  fidelity criterion 
 \begin{equation}\label{multifid2}
\bracket{\psi}{\rho}{\psi} \leq \max_{(P\lvert \bar{P})} \sum_{i=1}^{r} \lambda_i\left(\psi_P\right),
\end{equation}
for some target state $\ket{\psi}$, where the eigenvalues $\lambda_i(\psi_P)$ are ordered decreasingly. We will prove that if \eqref{eq:SDP_GME_dim2} is feasible then \eqref{multifid2} is satisfied. In other words, we will show that the SDP criterion is stronger at detecting the GME-dimension than is any possible fidelity criterion, regarless of how the target state is chosen.

To this end, we assume that $\{\tilde{\Pi}_{P}\}_{(P\lvert \bar{P})}$ is a feasible set for the SDP \eqref{eq:SDP_GME_dim2}. Since $\sum_{(P\lvert \bar{P})}\tilde{\Pi}_P\otimes \mathds{1}_{\bar{P}} -\rho \succeq 0$ it follows in particular that 
\begin{equation}\label{expr}
0\leq \bracket{\psi}{\sum_{(P\lvert \bar{P})}\tilde{\Pi}_P\otimes \mathds{1}_{\bar{P}} -\rho}{\psi} = \sum_{(P\lvert \bar{P})} \tr\left(\tilde{\Pi}_P\psi_P \right) - \bracket{\psi}{\rho}{\psi}.
\end{equation}
Let the spectral decompositions of $\tilde{\Pi}_P$ and $\psi_P$ be
\begin{align}
	& \tilde{\Pi}_P=\sum_{i=1}^{d_P} \mu_{i|P} \ketbra{\mu_{i|P}}{\mu_{i|P}}, \qquad \text{and} \qquad  \psi_P=\sum_{i=1}^{d_P}  \lambda_{i|P} \ketbra{\lambda_{i|P}}{\lambda_{i|P}},
\end{align}
where we have ordered both $\mu_{i|P}$ and  $\lambda_{i|P}$ decreasingly for every choice of $P$. From \eqref{eq:SDP_GME_dim2} it follows that
\begin{equation}\label{condsmu2}
0\leq \mu_{i|P} \leq \frac{1}{r}\sum_{i=1}^{d_P} \mu_{i|P} \qquad \text{and} \qquad  \sum_{(P|\bar{P})}\sum_{i=1}^{d_P}  \mu_{i|P}  \leq r.
\end{equation}
This leads to
\begin{equation}
\tr\left(\tilde{\Pi}_P \psi_P\right) = \sum_{i,j=1}^{d_P} \lambda_{i|P} \mu_{j|P} \left|\braket{\lambda_{i|P}}{\mu_{j|P}}\right|^2 = \sum_{i,j=1}^{d_P} \lambda_{i|P} \mu_{j|P} T_{ij}^{P},
\end{equation}
where $T_{ij}^{P}=\left|\braket{\lambda_{i|P}}{\mu_{j|P}}\right|^2$ is a doubly stochastic matrix for every $P$. We can bound the right-hand-side from above by optimising $T_{ij}^P$ independently for each choice of $P$. Since the matrix is doubly stochastic and the eigenvalues ordered decreasingly, the optimal choice is $T_{ij}^P=\delta_{ij}$. Hence,
\begin{equation}
\sum_{(P\lvert \bar{P})} \tr\left(\tilde{\Pi}_P \psi_P\right) \leq \sum_{(P\lvert \bar{P})} \sum_{i=1}^{d_P} \lambda_{i|P} \mu_{j|P}.
\end{equation}
For simplicity, we re-write the conditions \eqref{condsmu2} as
\begin{equation}\label{condsmu3}
0\leq \mu_{i|P} \leq t_P \qquad \text{and} \qquad  \sum_{(P|\bar{P})}  t_P  \leq 1,
\end{equation}
where we have defined $t_P=\frac{1}{r}\sum_{i=1}^{d_P} \mu_{i|P}$. Thus, since the eigenvalues are ordered decreasingly, it is for a given value of $t_P$ optimal to choose the first $r$ eigenvalues as equal to $t_P$, implying that the  remaining eigenvalues are zero, i.e.~
\begin{equation}
\mu_{i|P}=\begin{cases}
t_P & \text{for }i=1,\ldots,r\\
0 &\text{otherwise} 
\end{cases}.
\end{equation}

 This gives
\begin{equation}
\sum_{(P\lvert \bar{P})} \tr\left(\tilde{\Pi}_P \psi_P\right) \leq \sum_{(P\lvert \bar{P})}t_P \sum_{i=1}^{r} \lambda_{i|P} .
\end{equation}
Since $t_P$ is a probability distribution, the optimal choice is to place the full weight on the largest element in the set $\{\sum_{i=1}^{N_P} \lambda_{i|P}\}_P$. This gives
\begin{equation}
\sum_{(P\lvert \bar{P})} \tr\left(\tilde{\Pi}_P \psi_P\right) \leq \max_{(P\lvert \bar{P})} \sum_{i=1}^{r} \lambda_{i|P}.
\end{equation}
Inserting the right-hand-side into \eqref{expr} gives
\begin{equation}
0\leq \max_{(P\lvert \bar{P})} \sum_{i=1}^{r} \lambda_{i|P}- \bracket{\psi}{\rho}{\psi},
\end{equation}
which is the fidelity criterion.


\section{SDP relaxations for dimension-restricted prepare-and-measure scenarios}\label{app:dimension_hierarchy}
A natural problem to which one may apply BMM methodology is to bound the set of dimension-restricted  correlations in prepare-and-measure scenarios. Here, we show that the most straightforward approach to this problem leads to an SDP relaxation hierarchy equivalent to the one proposed in \cite{jef_almost_qudit}. 

Consider a prepare and measure scenario, where Alice selects an input $x=\{1,\dots, n\}$ and send the $d$-dimensional state $\rho_x$ to Bob. Bob selects $y=\{1, \dots, m \}$ and performs the measurement $M_{b|y}$ to obtain the outcome $b$. Our objective is to verify whether the observed correlations $p(b|x,y)=\tr(\rho_x M_{b|y})$ are compatible with a quantum realization of the experiment where also convexification via a shared classical variable is permitted. For this, we choose the map $\Theta(X)=\Pi X \Pi$, where $\Pi$ is the identity on the $d$-dimensional space where the states $\{\rho_x\}_x$ are supported.  We define the list of relevant operators in the problem $L=\{\rho_x, M_{b|y}\}$ and build the BMM $\Gamma$ in the standard way, as monomials over $L$. The blocks in the BMM have size $d$ and the reduction rules become $\Pi \rho_x=\rho_x$, $\rho^2_x=\rho_x$ (purity) and since a Naimark dilation always can be applied to the measurements we have $M_{b\lvert y}M_{b'\lvert y}= \delta_{b,b'}M_{b\lvert y}$. The SDP becomes  
\begin{equation}\label{eq:SDP_dimension_restricted}
	\begin{aligned}
		\text{find} & & & \Gamma \\
		\text{s. t.} & & &  p(b|x,y)=\tr(\Gamma_{\rho_x, M_{b|y}})\quad \forall ~x,b,y\\
		 & & &  \tr(\Gamma_{\Pi, \Pi})=d\\
		 & & &  \Gamma_{\rho_x, \rho_x}=\Gamma_{\Pi,\rho_x} \quad \forall x\\
		& & & \tr(\Gamma_{\rho_x})=1 \qquad \forall ~x\\ 
		 & & &  \Gamma_{M_{b|y}, M_{b'|y}}=\delta_{b,b'}\Gamma_{\Pi,M_{b|y}} \quad \forall~b,y\\		 		
		& & & \sum_b \Gamma_{\Pi,M_{b|y}} = \Gamma_{\Pi,\Pi} \quad \forall ~y \\
		& & & \Gamma\succeq 0.
	\end{aligned}
\end{equation}
However, one can eliminate the block structure without loss of generality. The observation that makes this possible is that the SDP is invariant under the action of the unitary group $\mathcal{U}(d)$. Specifically, consider any unitary $U\in\mathcal{U}(d)$ and apply it to the monomials. This gives
\begin{equation}
	\Gamma^U = \sum_{u,v} U \Pi u v^\dagger\Pi U^\dagger \otimes \ket{i_u}\bra{i_v} = U\otimes\mathds{1} \left( \sum_{u,v} \Pi uv^\dagger\Pi \otimes \ket{i_u}\bra{i_v} \right) U^\dagger\otimes\mathds{1}.
\end{equation}
All the constraints in the SDP \eqref{eq:SDP_dimension_restricted} are left unchanged if we replace $\Gamma$ with $\Gamma^U$. Moreover, since the feasible set of the problem is convex, we can consider the  BMM obtained from averaging over whole unitary group via the Haar measure $\mu(U)$. That gives us the BMM 
\begin{equation}
	\overline{\Gamma}=\int d\mu(U)\Gamma^U=\sum_{u,v}\left[\int d\mu(U)U \Pi u v^\dagger \Pi U^\dagger \right] \otimes \ket{i_u}\bra{i_v}=\sum_{u,v}\tr(\Pi uv^\dagger \Pi)\dfrac{\mathds{1}}{d}\otimes \ket{i_u}\bra{i_v},
\end{equation}
where we have used the  Haar twirl, namely $\int d\mu(U) UXU^\dagger= \frac{\tr(X)}{d}\mathds{1}$. This shows that all the blocks in the BMM can without loss of generality be taken proportional to the identity operator. On the level of the SDP, that is equivalent to considering the tracial moment matrix  $\Gamma=\sum_{u,v}\tr(\Pi uv^\dagger) \ket{i_u}\bra{i_v}$ which is proposed  in \cite{jef_almost_qudit}.

\section{SDP relaxations for high-dimensional quantum steering}\label{app:steering}

Quantum steering represents the modern formulation of the Einstein-Podolsky-Rosen paradox \cite{Wiseman2007}. In this quantum scenario, two parties, Alice and Bob, share an entangled state $\rho_{AB}$. Alice performs measurements $\{A_{a|x}\}_{a,x}$ with $a$ defining the outcome and $x$ the measurement setting. Bob aims to characterize the properties of $\rho_{AB}$ by analysing the remote probabilistic states, or assemblage, generated by Alice's measurement $\sigma_{a|x}=\tr_A(A_{a|x}\otimes\mathds{1}\rho_{AB})$. To apply the BMM methodology to such problems, it is natural to choose the completely positive map used in building the BMM as $\Theta(X_{AB})=\tr_A(X_{AB})$. This map discards the uncharacterised part of the global Hilbert space and leaves us with the characterised part, namely that beloning to Bob, on which the assemblage is defined. 

Genuinely high-dimensional steering was introduced in \cite{Designolle2021}. It aims to certify from the assemblage alone the Schmidt number of the underlying quantum state. The Schmidt number of a bipartite density matrix $\rho_\text{AB}$ is defined as 
\begin{equation}\nonumber
	r(\rho_\text{AB})\equiv \min_{\{q_i\},\{\psi_i\}}  \Big\{r_\text{max}: \quad  \rho_\text{AB}=\sum_i q_i \ketbra{\psi_i} \quad \text{and} \quad r_\text{max}=\max_i \rank(\psi_i^\text{A})\Big\} \ ,
\end{equation}
where $\psi_i^\text{A}=\tr_B \ketbra{\psi_i}$ is the reduction of the pure state $\psi_i$. 

We follow the outlined methodology to construct a BMM SDP relaxation for the problem of certifying the Schmidt number in steering tests. Without loss of generality we can consider the state to be pure, namely $\rho_{AB}=\ketbra{\psi_{AB}}$, and Bob's Hilbert space to have a given dimension $d$. If $\ket{\psi_{AB}}$ has Schmidt rank $r$, then it admits a Schmidt decomposition
\begin{equation}
	\ket{\psi}_{AB}=\sum_{j=1}^{r}\sqrt{\lambda_j}\,\ket{\alpha_j}_A\ket{\beta_j}_B,
	\qquad \lambda_j>0,\ \ \sum_{j=1}^r \lambda_j=1,
\end{equation}
where $\{\ket{\alpha_j}\}$ and $\{\beta_j\}$ respectively are orthonormal vectors. The reduced state becomes  $\rho_A=\tr_B(\ketbra{\psi})=\sum_{j=1}^r \lambda_j \ketbra{\alpha_j}$ and it is fully supported on a local Hilbert space of rank no larger than $r$. We call $\Pi=\sum_{j=1}^r \ketbra{\alpha_j}{\alpha_j}$ the projector onto this local support. Hence, states with Schmidt number no larger than $r$ satisfy 
\begin{align}\label{props}
	\Pi^2=\Pi,\quad \tr(\Pi)\leq r,\quad \text{and}\quad (\Pi\otimes \mathds{1}) \ket{\psi}_\text{AB}=\ket{\psi}_\text{AB}.
\end{align}
Moreover, since the state has no support locally outside $\Pi$,  we can also impose that the measurements satisfy $\Pi A_{a|x}=A_{a|x}$. 	

The relevant operators in the scenario are the shared state $\rho_{\text{AB}}=\ketbra{\psi}{\psi}$, the local uncharacterized measurements $\{A_{a|x}\}_{a,x}$ and the projector on the relevant local support $\Pi$. The minimal list of monomials is then $L=\{\Pi_A\otimes\mathds{1}, A_{a|x}\otimes\mathds{1}, \rho_{AB}\}$. After building the BMM corresponding to level $S$, we can enforce the constraints in \eqref{props} through the linear equality $\Gamma_{\Pi_A\otimes\mathds{1}, \Pi_A\otimes\mathds{1}}=\alpha\mathds{1}$, $\alpha\leq r$ and $\Gamma_{\Pi_A\otimes\mathds{1}, \psi_{AB}}=\rho_{B}$. Along with the definition of the assemblage $\Gamma_{A_{a|x}\otimes\mathds{1}, \psi_{AB}}=\sigma_{a|x}$ and the measurement support $\sum_a\Gamma_{A_{a|x}\otimes\mathds{1}, \Pi_A\otimes\mathds{1}}=\alpha\mathds{1}$ this yields an SDP that tests whether a given assemblage can arise from measurements on a pure state with Schmidt rank at most $r$. To clarify, consider the illustrative list of monomials $L=\{\Pi_A\otimes\mathds{1}, \rho_{\text{AB}}, A_{0|0}\otimes\mathds{1}, A_{0|1}\otimes\mathds{1}, A_{1|0}\otimes\mathds{1}\rho_{\text{AB}} \}$. Recalling that the blocks are obtained from the map $\Theta=\tr_A$ and using the aforementioned constraints, the BMM reads 
\begin{equation}\label{eq:app_C}
	\Gamma=\begin{pmatrix}
		\alpha\mathds{1} & \rho_B & a_{0|0}\mathds{1} & a_{0|1}\mathds{1} & \sigma_{1|0} \\[0.5ex]
		&  \rho_B & \sigma_{0|0}  & \sigma_{0|1} & Q_1 \\[0.5ex]
		& & b_{0,0}\mathds{1} & b_{0,1}\mathds{1} & 0\\[0.5ex]
		& & & b_{1,1}\mathds{1} & Q_2 \\
		& & & & Q_3
	\end{pmatrix},
\end{equation}
where $a_{a|x}$ and $b_{x,x'}$ represents scalars and $\{Q_i\}$ matrix variables. Additionally, $a_{a|x}\geq b_{x,x}$ follows from $\Gamma_{\Pi\otimes\mathds{1}, A_{a|x}\otimes\mathds{1}}=\tr(A_{a|x}\Pi)\mathds{1}=\tr(A_{a|x})\mathds{1}\geq\tr(A_{a|x}A_{a|x})\mathds{1}$ and $Q_1\succeq 0$ since $\tr_A(\rho_{\text{AB}}A_{1|0}\rho_{\text{AB}})\succeq 0$ while $Q_2$ is any matrix. Furthermore, since the feasible set is convex, the same criterion also applies to mixed states: if $\rho_{AB}=\sum_i q_i\ketbra{\psi_i}$ with each $\ket{\psi_i}$ of Schmidt rank $\le r$, then the corresponding matrices $\Gamma^{(i)}$ are feasible and so is their mixture $\Gamma=\sum_i q_i\Gamma^{(i)}$, reproducing the assemblage of $\rho_{AB}$. Considering for simplicity projective measurements, the SDP can then be written as 
\begin{equation}\label{eq:SDP_steering_dimension}
	\begin{aligned}
		\text{find} & & & \Gamma \\
		\text{s. t.} & & &  \sigma_{a|x}=\Gamma_{A_{a|x}\otimes\mathds{1}, \rho_\text{AB}}\quad \forall ~a,x\\
		& & & \sum_a \sigma_{a|x}=\rho_{B} \qquad \forall ~x\\ 
		& & &  \Gamma_{\rho_{\text{AB}}, \rho_{\text{AB}}}=\rho_{B} \\
		& & &  \tr(\Gamma_{\Pi\otimes\mathds{1}, \Pi\otimes\mathds{1}})=\alpha\mathds{1}, \quad \alpha\leq r\\
		& & &  \Gamma_{A_{a|x}\otimes\mathds{1}, A_{a'|x}\otimes\mathds{1}}=\delta_{a,a'}\Gamma_{\Pi\otimes\mathds{1},A_{a|x}\otimes\mathds{1}} \quad \forall~x\\		 		
		& & & \sum_a\Gamma_{A_{a|x}\otimes\mathds{1}, \Pi_A\otimes\mathds{1}}=\alpha\mathds{1} \quad \forall ~x \\
		& & & \Gamma\succeq 0.
	\end{aligned}
\end{equation}

This formulation mirrors precisely the SDP approach introduced in \cite{SDP_Steering}.

\twocolumngrid
\bibliography{bibliography.bib}
\end{document}